\documentclass[manuscript,screen]{acmart}

\usepackage{multirow}
\usepackage{booktabs}
\usepackage[most]{tcolorbox}
\usepackage{graphicx}
\usepackage{ulem}
\usepackage{subcaption}
\usepackage{bm}
\usepackage[ruled,vlined]{algorithm2e}

\definecolor{c3}{cmyk}{0.3081,0,0.7209,0.3255}

\newtcbox{\hlprimarytab}{on line, rounded corners, box align=base,
  colback=c3!10,colframe=white,size=fbox,arc=3pt,
  before upper=\strut, top=-2pt, bottom=-4pt,
  left=-2pt, right=-2pt, boxrule=0pt}

\newtcbox{\hlsecondarytab}{on line, box align=base,
  colback=red!10,colframe=white,size=fbox,arc=3pt,
  before upper=\strut, top=-2pt, bottom=-4pt,
  left=-2pt, right=-2pt, boxrule=0pt}

\newcommand{\dashifted}{\raisebox{0.5\depth}{\tiny$\downarrow$}}
\newcommand{\uashifted}{\raisebox{0.5\depth}{\tiny$\uparrow$}}
\newcommand{\da}[1]{{\scriptsize\hlsecondarytab{\dashifted#1}}}
\newcommand{\ua}[1]{{\scriptsize\hlprimarytab{\uashifted#1}}}

\newcommand{\df}[1]{{\scriptsize\hlprimarytab{\dashifted#1}}}

\definecolor{c3}{cmyk}{0.3081,0,0.7209,0.3255}

\newtcolorbox{SummaryBox}{
  enhanced,
  breakable,
  colback=blue!5,        % 淡蓝背景
  colframe=blue!40,      % 左边框颜色
  boxrule=0pt,           % 不要四周边框线
  leftrule=3pt,          % 左侧强调线
  rightrule=0pt,
  toprule=0pt,
  bottomrule=0pt,
  arc=3pt,               % 圆角
  left=1pt,
  right=1pt,
  top=1pt,
  bottom=1pt,
}
\AtBeginDocument{%
  }

\setcopyright{acmlicensed}
\copyrightyear{2018}
\acmYear{2018}
\acmDOI{XXXXXXX.XXXXXXX}
\acmConference[Conference acronym 'XX]{Make sure to enter the correct
  conference title from your rights confirmation email}{June 03--05,
  2018}{Woodstock, NY}
\acmISBN{978-1-4503-XXXX-X/2018/06}

\newcommand{\ourmodel}{\textsc{MRCoder}}
\begin{document}

\title{\ourmodel: An Efficient Context Selecting Approach for Repository-Level Code Generation}


\author{Peiding Wang}
\email{wangpeiding@buaa.edu.cn}
\affiliation{%
  \institution{Beihang University}
  \department{School of Computer Science \& Engineering, State Key Laboratory of Complex \& Critical Software Environment}
  \city{Beijing}
  \country{China}}

\author{Li Zhang}
\email{lily@buaa.edu.cn}
\affiliation{%
  \institution{Beihang University}
  \department{School of Computer Science \& Engineering, State Key Laboratory of Complex \& Critical Software Environment}
  \city{Beijing}
  \country{China}}

\author{Fang Liu}
\authornote{Corresponding author.}
\email{fangliu@buaa.edu.cn}
\affiliation{%
  \institution{Beihang University}
  \department{School of Computer Science \& Engineering, State Key Laboratory of Complex \& Critical Software Environment}
  \city{Beijing}
  \country{China}}

\renewcommand{\shortauthors}{Wang et al.}

\begin{abstract}
Large language models (LLMs) have demonstrated strong capabilities in code generation. However, repository-level code generation remains challenging, as it requires effectively identifying and utilizing repository-specific context. While retrieval-augmented generation (RAG) incorporates relevant code snippets, it often introduces redundant context that interferes with the LLM's ability to utilize relevant information, leading to degraded generation quality and increased computational cost. Moreover, existing context selection and compression methods struggle to balance efficiency and quality, either introducing additional computational overhead or failing to effectively select valid context.
In this paper, we propose \ourmodel{}, an efficient context selection framework that improves both the effectiveness and efficiency of repository-level code generation. \ourmodel{} adopts a Map–Reduce paradigm: in the \textit{}{Map Phase}, a lightweight draft model generates drafts over partitioned contexts, and Structure-Aware Draft-Guided Selection (SADGS) selects informative contexts based on drafts through API consistency and logical similarity; in the \textit{Reduce Phase}, the refined contexts are aggregated for final generation, with a parallel verification strategy further accelerating decoding.
We evaluate \ourmodel{} on two widely used repository-level code generation benchmarks, CoderEval and DevEval, using Qwen2.5-Coder and DeepSeek-Coder as backbone LLMs. 
Experimental results show that \ourmodel{} improves code generation accuracy over strong baselines while reducing token consumption by 30\%–50\% and inference time by up to 52\%. These results demonstrate that our proposed structured and draft-guided context selection strategy is crucial for improving both the quality and efficiency of repository-level code generation. The code and data are available at \href{https://github.com/zhu-zhu-ding/MRCoder}{https://github.com/zhu-zhu-ding/MRCoder}.
\end{abstract}





\maketitle

\section{Introduction}
Recent advances in large language models (LLMs) have fundamentally transformed the landscape of software development~\cite{codellama,codeifbench,Cursor}, enabling intelligent systems to assist developers throughout the entire programming lifecycle. The emergence of powerful code-centric LLMs, such as Qwen-Coder~\cite{qwen25coder} and DeepSeek-Coder~\cite{deepseek}, has substantially enhanced the ability of AI systems to infer developer intent and generate high-quality code, thereby improving development efficiency~\cite{efficientedit}. While early studies primarily focused on isolated function-level generation~\cite{mbpp,humaneval}, recent research has increasingly shifted toward the more realistic and challenging setting of repository-level code generation~\cite{crosscodeeval,codereval}. In this scenario, models must leverage rich repository-level context, including cross-file dependencies, user-defined APIs, and project-specific conventions, to produce functionally correct and contextually consistent code~\cite{deveval}. 

% To address these challenges, Retrieval-Augmented Generation (RAG) has become the dominant paradigm for incorporating repository-level context into code generation~\cite{repocoder,rl-coder}. By retrieving relevant code snippets—such as definitions, usage patterns, and dependency-related components—from across the repository and integrating them into the model input, RAG-based approaches effectively extend the model’s contextual awareness and improve generation accuracy. For example, RepoCoder~\cite{repocoder} introduces an iterative retrieval and generation process that retrieves more relevant code snippets by optimising queries in subsequent rounds, while RLCoder~\cite{rl-coder} employs a reinforcement learning framework to train the retriever itself, enabling it to retrieve more relevant code snippets.
Retrieval-Augmented Generation (RAG) has become the dominant paradigm for incorporating repository-level context into code generation~\cite{repocoder,rl-coder}. By retrieving relevant code snippets—such as definitions, usage patterns, and dependencies—and integrating them into the model input, RAG enhances contextual awareness and improves generation accuracy. For instance, RepoCoder~\cite{repocoder} adopts an iterative retrieval-generation framework that refines queries across rounds, while RLCoder~\cite{rl-coder} leverages reinforcement learning to optimize the retriever for more effective context selection. 
Despite these advances, existing RAG-based methods share a fundamental limitation:
\textbf{not all retrieved code contexts are truly relevant to the generation process}.
In practice, retrieval inevitably introduces noisy or weakly relevant contexts, especially when the number of retrieved chunks is increased to improve recall~\cite{ragnoise,crag,repoformer}.
The presence of such noisy contexts leads to two major issues:

\textbf{Issue 1: Degraded Generation Quality}. Irrelevant or redundant contexts can distract the LLM and introduce conflicting signals, causing incorrect reasoning, hallucinations, or off-target code generation. Moreover, noisy contexts unnecessarily increase the input length to make LLMs struggle to effectively utilize all provided information, resulting in diminished accuracy.

\textbf{Issue 2: Reduced Generation Efficiency.} Noisy contexts unnecessarily increase the input length, thereby raising the computational cost of LLM inference and slowing down generation. In repository- or codebase-level code generation scenarios, such increased latency can significantly hinder developer productivity, resulting in a poor user experience~\cite{fastcoder}.

\begin{figure}[t]
\centering
\setlength{\abovecaptionskip}{0.1cm}

\begin{subfigure}[t]{0.49\linewidth}
    \centering
    \includegraphics[width=\linewidth]{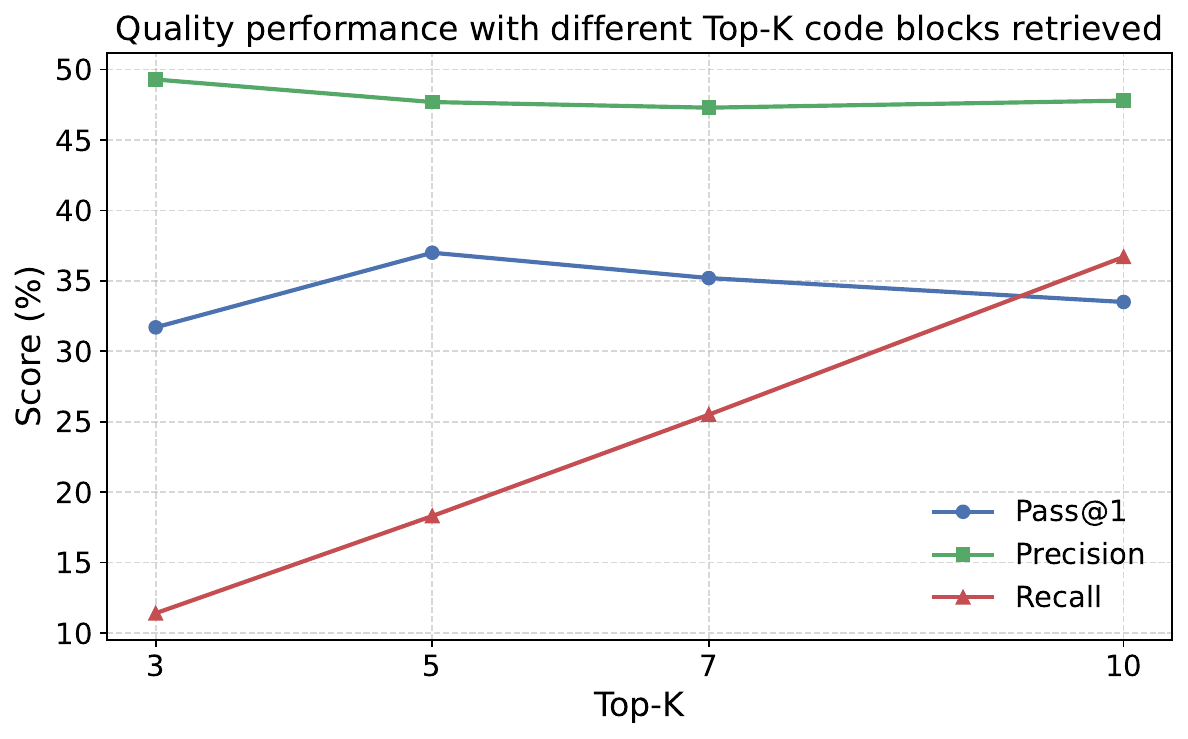}
    \label{fig:perf}
\end{subfigure}
\hfill
\begin{subfigure}[t]{0.49\linewidth}
    \centering
    \includegraphics[width=\linewidth]{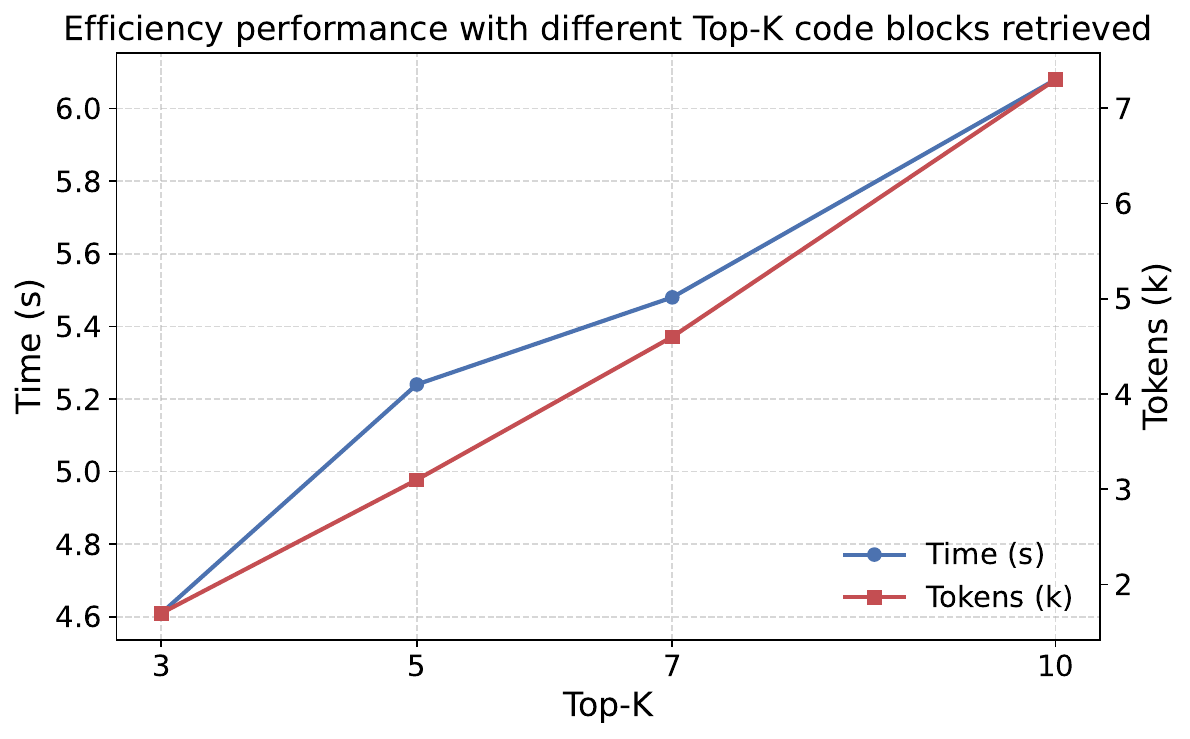}
    \label{fig:eff}
\end{subfigure}

\caption{\textbf{Motivation Example}. Results from Qwen2.5-Coder-7B-Instruct on CoderEval with BM25 retrieval. As Top-$K$ increases, recall improves, but redundant and noisy context degrades generation quality after an initial gain. Meanwhile, longer contexts increase both token consumption and generation time, leading to reduced efficiency.}
\label{fig:motivation}
\end{figure}

As illustrated in Figure \ref{fig:motivation}, as the amount of effective context increases (i.e., recall improves), the generation quality (Pass@1) initially improves but subsequently declines. At the same time, the steady drop in precision suggests that redundant and noisy contexts increasingly interfere with the LLM’s ability to leverage truly relevant information. In addition, these noisy contexts inflate the input length, leading to higher inference costs and longer generation latency. Together, these observations highlight a key challenge in repository-level code generation: effectively identifying and selecting useful context while filtering out.

Context selection and compression have been explored as potential solutions. RepoFormer~\cite{repoformer} and CodeFilter~\cite{codefilter} trains a small LLM to output control tokens for deciding whether to retrieve or select relevant contexts; however, it does not fundamentally mitigate the negative effects of retrieval noise (Issues 1 and 2). In contrast, LongCodeZip~\cite{longcodezip} reduces context length by selecting informative code lines based on perplexity computed by a small LLM, but requires multiple forward passes, leading to significant time overhead and limiting its applicability in latency-sensitive scenarios (Issue 2).
To address the above challenges, we propose \ourmodel{}, an efficient context selection framework for repository-level code generation. \ourmodel{} is designed to identify and retain code contexts that are truly beneficial for generation from a large pool of initially retrieved candidates. The framework consists of two main phases:

\textbf{Phase I: Map Phase.} We first partition the retrieved code blocks into multiple context groups and combine each group with the target code query to construct a set of prompt sequences. These prompts are then fed into a smaller LLM to generate code drafts in parallel. Based on these drafts, we apply a selection strategy that extracts relevant contexts from the candidate pool using signals such as API call consistency and semantic similarity. This process enables the smaller LLM
to evaluate context usefulness under moderate input lengths, effectively filtering out noisy or irrelevant code fragments.

\textbf{Phase II: Reduce Phase.} In this phase, the filtered contexts produced in the Map phase are aggregated and provided to a larger target LLM for final code generation. By focusing on high-quality and concise context, this step not only improves generation accuracy but also reduces input length, allowing the model to better utilize the most relevant information. In addition, we leverage the drafts generated during the Map phase to design a parallel verification mechanism for the target LLM. By validating multiple draft candidates concurrently, our approach accelerates the decoding process and reduces overall inference latency.

Overall, \ourmodel{} improves generation quality by selecting informative context fragments while simultaneously enhancing efficiency through context compression and draft-based parallel verification.

% To evaluate the effectiveness of \ourmodel{}, we conduct extensive experiments on two widely used repository-level code generation benchmarks, CoderEval~\cite{codereval} and DevEval~\cite{deveval}. We compare our approach against representative baselines, including standard RAG, the advanced repository-level code generation method RL-Coder~\cite{rl-coder}, and state-of-the-art code context compression approaches such as RepoFormer~\cite{repoformer} and LongCodeZip~\cite{longcodezip}. The evaluation is conducted across advanced open-source LLM families, namely Qwen2.5-Coder and DeepSeek-Coder. Experimental results show that \ourmodel{} consistently improves code generation accuracy while reducing computational cost. \textcolor{red}{Specifically, it outperforms all baseline methods in most settings, achieving up to \textbf{52.7\%} relative improvement in Pass@1 on CoderEval and \textbf{28.7\%} on DevEval compared to RAG, while maintaining more stable performance across different values of $K$.
% In terms of efficiency, \ourmodel{} reduces token consumption by \textbf{30\%--50\%} and decreases total inference time by up to \textbf{52.1\%} on CoderEval and \textbf{48.6\%} on DevEval compared to RAG, achieving a better balance between effectiveness and efficiency than existing methods.}
To evaluate the effectiveness of \ourmodel{}, we conduct extensive experiments on two widely used repository-level code generation benchmarks, CoderEval~\cite{codereval} and DevEval~\cite{deveval}. We compare our approach against representative baselines, including standard RAG, the advanced repository-level code generation method RL-Coder~\cite{rl-coder}, and state-of-the-art code context compression approaches RepoFormer~\cite{repoformer} and LongCodeZip~\cite{longcodezip}. The evaluation is conducted across advanced code LLM families, namely Qwen2.5-Coder and DeepSeek-Coder. Experimental results show that \ourmodel{} consistently improves code generation accuracy while reducing computational cost. Specifically, it outperforms all baseline methods in most settings, achieving up to \textbf{52.7\%} relative improvement in Pass@1 compared to RAG and \textbf{31.3\%} compared to state-of-the-art baseline LongCodeZip, while maintaining more stable performance across different retrieval blocks.
In terms of efficiency, \ourmodel{} reduces token consumption by \textbf{30\%--50\%} and decreases total inference time by up to \textbf{52.1\%} compared to RAG and \textbf{47.6\%} compared to state-of-the-art baseline RepoFormer, achieving a better balance between effectiveness and efficiency than existing methods.

Our contributions can be summarized as follows:
\begin{itemize}
\item We propose \ourmodel{}, a novel Map–Reduce framework for repository-level code generation that performs context selection after retrieval and before generation, effectively improving the quality and efficiency of generation.
\item We introduce Structure-Aware Draft-Guided Selection (SADGS), which leverages API call relationships and logical similarity between draft code and candidate contexts to identify informative context fragments and improve code quality.
% \item \textcolor{red}{We design a draft-based parallel verification mechanism that accelerates autoregressive decoding while preserving generation quality.}
\item We propose a draft-based efficiency optimization mechanism that leverages code drafts to select relevant contexts, reducing input length, and employs parallel verification to improve generation efficiency.
\item We conduct comprehensive experiments on two repository-level code generation benchmarks using different backbone LLMs. The results demonstrate that \ourmodel{} outperforms state-of-the-art context selection and compression methods, achieving improvements in both generation quality and efficiency.
\end{itemize}

% \begin{center}
% \textit{Are all retrieved code contexts useful for LLMs?}
% \end{center}

\section{Related Work}
\textbf{Repository-Level Code Generation.} Repository-level code generation has recently attracted increasing attention as large language models (LLMs) are applied to realistic software engineering scenarios~\cite{deveval,codereval,repoexec,crosscodeeval}. Unlike function-level benchmarks such as HumanEval~\cite{humaneval} and MBPP~\cite{mbpp}, repository-level tasks require LLMs to reason over long-range dependencies, cross-file interactions, and project-specific APIs.
A common paradigm for repository-level code generation is retrieval-augmented generation (RAG)~\cite{codeagent,rl-coder,repocoder}, where relevant code snippets are retrieved from the repository and provided as additional context to the LLM. Early work such as RepoCoder~\cite{repocoder} demonstrates that retrieving cross-file contexts can significantly improve code completion performance by exposing the LLM to relevant APIs and usage patterns. Similarly, GraphCoder~\cite{graphcoder} incorporate structural signals such as call graphs or dependency graphs to enhance retrieval quality. RL-Coder~\cite{rl-coder} improves the quality of generated code by using reinforcement learning to train a retriever to retrieve more relevant code. These approaches highlight the importance of grounding generation in repository-specific knowledge. However, they typically rely on Top-$K$  retrieval, which inevitably introduces substantial noise particularly when Top-$K$  is increased to improve retrieval recall~\cite{repoformer,codefilter}, leading to degraded model performance due to context overload.

\textbf{Code Context Selection and Filtering.} To select and retain the relevant code context, recent work has explored both implicit and explicit strategies to address this challenge. Repoformer~\cite{repoformer} proposes a selective retrieval framework that enables the model to decide whether retrieval is necessary, thereby avoiding unnecessary or harmful context and improving both efficiency and robustness. Similarly, CodeFilter~\cite{codefilter} employs a likelihood-based metric to train the model to assess the impact of retrieved context blocks on generation quality, thereby enabling the selection of beneficial code context. Complementarily, DietCode~\cite{DietCode} combines frequency-based filtering with CodeBERT~\cite{codebert} attention heuristics to remove low-impact tokens and SlimCode~\cite{Slimcode} employs rule-based token pruning using token types and program dependency graphs, which may generalize poorly across languages and tasks. However, both methods primarily focus on single-function compression for short-context scenarios. To address this, LongCodeZip~\cite{longcodezip} introduces a code-specific compression framework that performs hierarchical filtering at both function and block levels using perplexity-based relevance scoring, achieving substantial context reduction while preserving task performance.

Despite these advances, existing methods struggle to balance generation quality and efficiency. While Repoformer or CodeFilter use a lightweight model to generate a single decision token for retrieval control, it cannot effectively mitigate noisy retrieval. In contrast, LongCodeZip relies on multiple rounds of forward passes to compute perplexity for code selection, incurring substantial computational overhead.

\section{Method}
\subsection{Task Definition}
\label{sec:task_definition}
% In this section, we formalize the task of repository-level code generation. 
% Let ${R} = \{c_1, \dots, c_N\}$ denote a code repository, where each $c_i$ represents a code context, such as a function or a class. 
% Given a query $q$, which consists of a prompt with , the objective is to let LLM generate the target function $f$ by leveraging the Top-$K$ retrieved contexts $\{c_1, \dots, c_k\}$. 
% The generation process can be formulated as:
% \begin{equation}
% f = \mathrm{LLM}(q,\{c_1, \dots, c_k\})
% \end{equation}
% In this work, our goal is to select contexts from the Top-$K$ that are are effective for generation process, thereby reducing irrelevant contexts and improving both the quality and efficiency of code generation.

In this section, we formalize the task of repository-level code generation~\cite{deveval,codereval,fastcoder}.
Given a code repository, we focus exclusively on its source code and abstract it into a collection of code contexts obtained via static analysis tools. Specifically, we parse the repository and decompose it into a set of semantically meaningful code units (functions and classes). Each such unit is treated as an individual code context. Formally, we denote this collection as $ \{C_1, \dots, C_N\}$, where each $C_i$ represents a code snippet corresponding to a function or class extracted from the repository.

Given a query $Q$, which consists of a prompt with code to be generated, the objective is to let LLM generate the target function $F$ by leveraging the Top-$K$ retrieved contexts $\{C_1, \dots, C_K\}$, where the Top-$K$ contexts correspond to the $K$ code snippets with the highest similarity to the query. 
The generation process can be formulated as:
\begin{equation}
F = \mathrm{LLM}(Q,\{C_1, \dots, C_K\})
\end{equation}
In this work, our goal is to select contexts from the Top-$K$ that are are effective for generation process, thereby reducing irrelevant contexts and improving both the quality and efficiency of code generation.

\subsection{Overview}
\begin{figure*}
	\centering
	\includegraphics[width=1\textwidth]{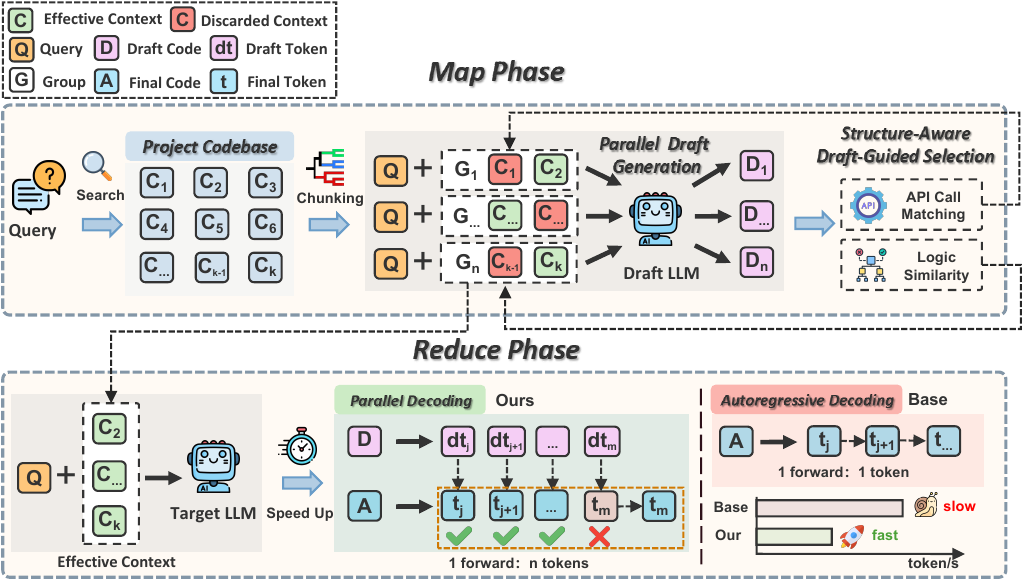}
 	\caption{The overall pipeline of \ourmodel{}. }
	\label{fig:total_process}
\end{figure*}

In this section, we introduce \ourmodel{}, a two-stage framework that selects effective contexts after retrieval and before generation. As shown in Figure \ref{fig:total_process}, given a query and a set of retrieved code contexts from repository, \ourmodel{} follows a Map--Reduce paradigm. In the \textit{Map Phase}, the retrieved contexts are partitioned into multiple groups, and a lightweight draft model generates draft code for each group independently. These drafts are then used to evaluate the usefulness of contexts via \textit{Structure-Aware Draft-Guided Selection (SADGS)} based on API consistency and semantic similarity, producing a filtered set of high-quality contexts. In the \textit{Reduce Phase}, the selected contexts are aggregated and fed into a larger target LLM for final code generation. Meanwhile, the intermediate drafts are further leveraged for parallel verification to accelerate decoding. 

% By decoupling context evaluation from final generation, \ourmodel{} effectively filters noisy contexts and reduces input length, leading to improved generation quality and efficiency.

\subsection{Map Phase}
\subsubsection{Context Partition and Draft Generation}
As described in Section \ref{sec:task_definition}, we first employ static analysis to extract a set of code contexts from the repository. Specifically, we use tree-sitter\footnote{\url{https://github.com/tree-sitter/tree-sitter}} to parse each source file into functions and classes code units. To mitigate potential data leakage, we filter out code contexts whose Jaccard Similarity~\cite{jaccardsimilarity} with the standard answer exceeds 0.9~\cite{joyai,SplitGuard}, thereby removing near-duplicate or highly similar code snippets. After this preprocessing step, we obtain the final code contexts $\{C_1, \dots, C_K\}$.
As shown in Figure \ref{fig:total_process}, following prior work~\cite{repocoder,rl-coder}, given a code generation query $Q$, we first retrieve the Top-$K$ relevant code contexts from $\{C_1, \dots, C_N\}$:
\begin{equation}
\{C_1, \dots, C_K\} = \mathcal{F}_{\text{retrieval}}(Q, \{C_1, \dots, C_N\})
\end{equation}
where $\mathcal{F}_{\text{retrieval}}(\cdot)$ denotes the retrieval method. Notably, the retrieval component is not tightly coupled with any specific approach; \ourmodel{} does not rely on particular retrieval optimizations and can be seamlessly integrated with existing retrieval methods.

Next, we partition the retrieved code contexts into multiple subgroups of size $M$, where $M$ is a predefined hyperparameter. As shown in Figure \ref{fig:total_process}, the $M$ is set to 2. Given the retrieved contexts $\{C_1, \dots, C_K\}$, we first rank them in descending order according to their similarity to the query $Q$. We then divide the ranked list into $n$ consecutive groups with a fixed stride of $M$:
\begin{equation}
\{C_1, \dots, C_K\} = \bigsqcup_{i=1}^{n} G_i, \quad |G_i| = M \ (i < n), \quad |G_n| \leq M
\end{equation}
where $n = \left\lceil \frac{K}{M} \right\rceil$, and each group $G_i$ contains a contiguous segment of contexts from the ranked list.
We also explored dynamic partitioning strategies based on clustering methods (e.g., Random and K-Means~\cite{kmeans}). However, these approaches did not yield noticeable performance gains while introducing additional computational overhead (See Section \ref{diss:partition_strategy}). Therefore, we adopt a simple and efficient fixed partitioning strategy.
% Next, we partition the retrieved code contexts into multiple subgroups of size $m$, where $m$ is a predefined hyperparameter. Formally, let $\{c_1, \dots, c_k\}$ denote the retrieved contexts. We divide them into $n$ groups:
% \begin{equation}
% \{c_1, \dots, c_k\} = \bigsqcup_{i=1}^{n} G_i, \quad |G_i| = m \ (i < n), \quad |G_n| \leq m
% \end{equation}
% where $n = \left\lceil \frac{k}{m} \right\rceil$.
% We also explored dynamic partitioning strategies based on clustering methods (e.g., K-Means~\cite{kmeans}). However, these approaches did not yield noticeable performance gains while introducing additional computational overhead. Therefore, we adopt a simple and efficient fixed partitioning strategy.

Then we combine each context group $G_i$ with the query $Q$ to construct independent prompts, which are processed in parallel by a lightweight draft model to obtain draft code sequences $\{D_1,\dots,D_n\}$. These drafts serve as intermediate signals that reflect how each context group contributes to the generation, and are subsequently used to guide context selection. 
Formally, for each group $G_i$, we define:
\begin{equation}
D_i = \mathrm{LLM}(Q, G_i), \quad i = 1, \dots, n
\end{equation}
where $D_i$ is the draft code generated by the small draft LLM corresponding to $G_i$. 
This design ensures that each prompt has a moderate context length, enabling the LLM to better utilize relevant information compared to directly processing the full set of retrieved contexts. Moreover, the parallel generation of drafts significantly reduces inference latency in the \textit{Map Phase}. 
\subsubsection{Structure-Aware Draft-Guided Selection}
In this section, we introduce Structure-Aware Draft-Guided Selection (SADGS), a dual-perspective context selection framework that leverages structural signals in code. Specifically, given the draft set $\{D_1, \dots, D_n\}$ generated in the \textit{Map Phase} and the retrieved contexts groups $\{G_1, \dots, G_n\}$, SADGS selects a subset of informative contexts ${G}^* \subseteq {G}$ based on two complementary signals. The design of SADGS is motivated by two key observations in repository-level code generation: (1) \textbf{API Call Perspective:} target implementations often rely on invoking repository-defined APIs, and (2) \textbf{Logic Perspective:} similar logic patterns in the repository can provide valuable references.

\textbf{API Call Perspective.}
For this perspective, our goal is to select valid contexts based on the API calls between the generated draft code and code contexts. We employ the static analysis tool tree-sitter to extract API call information from both the draft code and the retrieved contexts. In this paper, we categorize APIs into two types based on call relationships:
\begin{itemize}
\item \textbf{External APIs}: function or class calls that are invoked within the current code unit but defined outside of it, representing outgoing dependencies. These APIs reflect how the current code relies on external callable interfaces.

\item \textbf{Internal APIs}: functions or classes defined within the current code unit that can be invoked by other code units, representing exposed callable interfaces. These APIs characterize the reusable interfaces provided by the code unit itself.
\end{itemize}
For example, consider a class $A$ with a method $B$, where $B$ calls a function $C$ that is defined outside of class $A$. In this case, $C$ is treated as an external API of $B$, as it is defined outside the method but invoked within it. Conversely, $B$ is regarded as an internal API of class $A$, since it is defined within the class and can be called by other code units.

For each draft $D_i$, we identify its external API call set $\mathcal{A}(D_i)_{\text{ext}}$. This API sequence contains the necessary external API information from code context required to complete the draft code.
For each code context $C \in G_i$, we extract both its external API calls and internal API calls $\mathcal{A}(C_i)$. Internal API calls within the code context directly provide callable external information for the generated code. The reason why we retain external API calls in the context, as identical external API usage is that it can serve as reference patterns to guide the generation of code using this API. Based on above, we select context blocks that share overlapping API calls with the draft. Formally, the selected context subset is then defined as:
\begin{equation}
{G_i}^{API} = \{\, C_t \in {G_i} \mid \mathcal{A}(C_t) \cap \mathcal{A}(D_{i})_{\text{ext}} \neq \emptyset \,\},t \leq M
\end{equation}

This mechanism preserves both repository-specific APIs that the target implementation may invoke and reference code patterns for external API usage, thereby providing efficient and informative context for subsequent LLM-based generation from the API perspective.

% We employ the static analysis tool tree-sitter to extract API call information from both the draft code and the retrieved contexts. Specifically, for each draft $d_i$, we identify its external API call set $\mathcal{A}(d_i)_{\text{ext}}$. For each code context $c \in G_i$, we extract both its external API calls and internal API calls $\mathcal{A}(c_i)$, where external APIs are methods that call others, while internal APIs can be called by others. The reason why we retain external API calls in the context, as identical external API usage is that it can serve as reference patterns, providing guidance for code generation. Based on this intuition, we select context blocks that share overlapping external API calls with the draft. Formally, the selected context subset is then defined as:
% \begin{equation}
% {G_i}^{API} = \{\, c_t \in {G_i} \mid \mathcal{A}(c_t) \cap \mathcal{A}(d_{i})_{\text{ext}} \neq \emptyset \,\},t \leq m
% \end{equation}

% This mechanism preserves both repository-specific APIs that the target implementation may invoke and reference code patterns for external API usage, thereby providing efficient and informative context for subsequent LLM-based generation from the API perspective.

\textbf{Logic Similarity Perspective.}
To further capture logical consistency between the draft and candidate contexts, we perform similarity-based selection. The goal of this step is to identify code contexts that are most semantically and structurally aligned with the draft implementation. Specifically, for each draft $D_i$ and each context $C_t \in G_i$, we compute their relevance score using the BM25~\cite{BM25} similarity function:
\begin{equation}
S_{\text{sim}}(D_i, C_t ) = \text{BM25}(D_i, C_t )
\end{equation}
where $\text{BM25}(\cdot, \cdot)$ measures the logic and structural similarity between the draft code and the candidate context.

% Based on the computed similarity scores, we retain the Top-$l$ ($l$ is a hyperparameter and $l<m$) most relevant contexts:
% \begin{equation}
% G_i^{\text{sim}} = \operatorname{Top}_l \left( \{\, c_t  \in G_i \mid S_{\text{sim}}(d_i, c_t ) \,\} \right)
% \end{equation}
Based on the computed similarity scores, we rank all contexts $C_t  \in G_i$ in descending order according to $S_{\text{sim}}(D_i, C_t )$, and retain the Top-$L$ ($L < M$) most relevant ones:
\begin{equation}
G_i^{\text{SIM}} =
\left\{
C_t'\in \operatorname*{arg\,top}_{C_t \in G_i}^{\,L}
S_{\text{sim}}(D_i, C_t)
\right\}
\end{equation}
This mechanism preserves code contexts that exhibit similar implementation logic to the draft, thereby providing the subsequent LLM with concise and informative context from a logical perspective, complementary to the API-based selection.

Finally, we merge $G_i^{\text{API}}$ and $G_i^{\text{SIM}}$ and remove duplicate elements to obtain the final informative contexts $G^*$:
\begin{equation}
G_i^* = G_i^{\text{API}} \cup G_i^{\text{SIM}}.
\end{equation}
% We extract API calls from both drafts and code contexts using a static analysis tool (e.g., Tree-Sitter). Let $\mathcal{A}(d)$ and $\mathcal{A}(c)$ denote the sets of API calls in draft $d$ and context $c$, respectively. We retain contexts that share API calls with drafts:
% \[
% \mathcal{C}_{\text{API}} = \left\{ c \in \mathcal{C}_k \ \middle|\ \exists d \in \mathcal{D}, \ \mathcal{A}(d) \cap \mathcal{A}(c) \neq \emptyset \right\}.
% \]

% \textbf{Similarity-based selection.}  
% To capture logical relevance, we compute a similarity score between each context and the draft set using a lightweight BM25-based function $\mathrm{Sim}(\cdot)$. We select the top-$t$ most similar contexts:
% \[
% \mathcal{C}_{\text{sim}} = \operatorname{Top}_t \left( \{c \in \mathcal{C}_k\}, \ \mathrm{Sim}(c, \mathcal{D}) \right).
% \]

% Finally, the selected context set is obtained by combining both signals:
% \[
% \mathcal{C}^* = \mathcal{C}_{\text{API}} \cup \mathcal{C}_{\text{sim}}.
% \]

% Through this dual-perspective selection, SADGS effectively filters out noisy contexts while preserving structurally and semantically relevant information, providing high-quality inputs for the subsequent generation stage. Moreover, the drafts $\mathcal{D}$ also serve as high-quality intermediate outputs that can be further leveraged to accelerate the final decoding process.

\subsection{Reduce Phase}
Although the \textit{Map Phase} effectively filters out irrelevant context, the target LLM is still required to generate the final code in a token-by-token, autoregressive manner based on the refined context, which inherently limits generation efficiency. Moreover, the incorporation of draft models for parallel generation in \textit{Map Phase} introduces additional computational overhead. Therefore, in this phase, our primary objective is to \textbf{improve the efficiency of final code generation}.

\subsubsection{Preliminaries}

\textbf{Autoregressive decoding~\cite{vaswani2023attentionneed}} is a prevalent paradigm in LLM inference, where output tokens are generated sequentially through iterative forward predictions. 
Formally, given an LLM $\texttt{LLM}$, a prefix token sequence $(t_1, \dots, t_{n-1})$, and the current input token $t_n$, the model predicts the probability distribution over the next token as:
\begin{equation}
\bm{p}_{n+1} = \texttt{LLM}(t_n \mid t_1, \dots, t_{n-1})
\end{equation}
where $\bm{p}_{n+1}$ denotes the probability distribution for the next token.
The next token $t_{n+1}$ is then obtained either by selecting the highest probability token from $\bm{p}_{n+1}$ or by sampling according to the distribution. The selected token is subsequently appended to the sequence and fed back into the model to generate the next token. Inherently, this token-by-token nature of autoregressive decoding introduces computational inefficiencies, resulting in inference latency that scales linearly with both the generated sequence length and the model complexity.

\textbf{Parallel decoding~\cite{speculative}} aims to accelerate the autoregressive generation process of LLMs. While LLMs inherently generate one token per forward pass, their inference efficiency can be significantly improved by verifying and accepting multiple tokens in parallel. The target LLM accelerates the generation process by performing parallel validation and accepting token sequences from the draft sequence that match the final answer it has generated. Given a sequence of draft tokens generated by the small draft LLM in \textit{Map Phrase}, the target LLM generating the final answer can evaluate the entire sequence in a single forward and determine which tokens are consistent with its own predicted distributions. This mechanism allows the LLM to accept multiple tokens simultaneously, thereby reducing the number of required forward passes and accelerating the overall generation process. 

Formally, given a sequence of $m$ draft tokens $(dt_1, \cdots, dt_m)$, the target model is conditioned on the prefix $(t_1, \dots, t_n)$.
\begin{equation}
    \bm{p}_{1}, \cdots, \bm{p}_{m} = \texttt{LLM}_{\text{target}}(dt_1, \cdots, dt_m \mid t_1, \dots, t_n),
\end{equation}
where $\bm{p}_i$ denotes the predicted probability distribution for the $i$-th draft token position.
Each draft token $dt_i$ is then verified against the corresponding distribution $\bm{p}_i$. The generation token can be defined as:
\begin{equation}
    \begin{cases}
        dt_i, & \text{if } dt_i = decode( \bm{p}_i) \\
        decode( \bm{p}_i), & \text{if } dt_i \neq decode( \bm{p}_i)
    \end{cases}
\end{equation}
A draft token $dt_i$ is defined as \textbf{accepted} if it matches $\operatorname{decode}(\bm{p}_i)$, and is otherwise \textbf{rejected}. If a draft token $dt_j$ is rejected at position $j$ (where $j < m$), the verification process terminates immediately, and the token $t_j$ is instead sampled from the corresponding probability distribution $\bm{p}_j$. No further draft tokens are validated beyond this point; If all $m$ draft tokens are accepted, we set $j = m + 1$ and sample the next token $t_j$ from the distribution $\bm{p}_{m+1}$. Then the resulting token sequence $\{(dt_1, \dots, dt_{j-1},t_j),j\leq m+1\}$ generated in a single forward.

This parallel verification mechanism not only accelerates the autoregressive generation process of LLMs but also ensures that the final output strictly adheres to the model’s intrinsic probability distribution. As a result, generation quality is preserved when the same decoding or sampling strategy is employed.

\subsubsection{Phrase Pipline}
\begin{algorithm}[t]
\small
\caption{Efficient Parallel Decoding for Reduce}
\label{alg:reduce}
\KwIn{Query $Q$, refined context $G^*$, draft code tokens ${dt_1,, \dots, dt_m}$}
\KwOut{Generated token sequence $T$}

Initialize $T \leftarrow \emptyset$; \\
Set pointer $j \leftarrow 1$; \\

\While{len($T$) < MAX\_TOKENS}{
    \If{$j \leq m$}{
        \tcp{Parallel verification}
        $\bm{p}_j, \dots, \bm{p}_m \leftarrow \texttt{LLM}_{\text{target}}(dt_j, \dots, dt_m \mid q,G^*)$; \\

        Find the longest accepted prefix:
        \begin{equation*}
        k = \max \left\{ z \;\middle|\; dt_{j+i-1} = decode( \bm{p}_{j+i-1}), \; \forall i \in [1,z] \right\};
        \end{equation*}

        Accept $(dt_j, \dots, dt_{j+k-1})$; \\
        $T \leftarrow T \oplus (dt_j, \dots, dt_{j+k-1})$; \\
        $j \leftarrow j + k$; \\
        $t=decode( \bm{p}_j)$; \\
        $T \leftarrow T \oplus t$; \\
        \tcp{Span alignment with remaining draft}
        Find $s$ such that:
        \begin{equation*}
        (t_1, \dots, t_x) = (dt_s, \dots, dt_{s+x-1}), \quad s \in [j+1, m-x+1];
        \end{equation*}

        \If{such $s$ exists}{
            $j \leftarrow s + x$; \\
        }
        \Else{
            \tcp{Continue autoregressive decoding}
            $j \leftarrow m + 1$;
        }
        }
        \Else{
            \tcp{Autoregressive decoding}
            $\bm{p} \leftarrow \texttt{LLM}_{\text{target}}(\cdot \mid Q, G^*, T)$; \\
            $t=decode( \bm{p})$; \\
            $T \leftarrow T \oplus t$; \\
        }
    
        \If{last token in $T$ is EOS}{
            \textbf{break};
        }
}
\Return $T$;
\end{algorithm}
In this section, we describe the overall pipline of the \textit{Reduce Phase}. Given the refined context $G^*$ and the draft sequences obtained from the \textit{Map Phase}, we construct a new instruction by combining $G^*$ with the query $q$. This instruction is then fed into a larger target LLM to generate the final output. This design provides the target LLM with high-quality, multi-perspective code context, enabling more accurate and robust code generation. At the same time, by filtering out noisy or irrelevant information, it improves the model’s ability to effectively utilize relevant contextual signals. Next, we select the first draft code $D_0$ from the drafts obtained by the draft LLM as the candidate $(dt_1, \cdots, dt_m)$ for verification to accelerate the generation process of the target LLM, where $dt_i$ represents the $i$-th token in the draft sequence. This choice is motivated by two considerations: (i) \textbf{Draft quality:} the first draft group $G_1$ exhibits the highest semantic similarity to the query, making it the most likely to yield a high-quality solution; (ii) \textbf{Efficiency:} prioritizing the most probable candidate minimizes the verification overhead for the target LLM.

As shown in Algorithm \ref{alg:reduce}, during verification, if the target LLM rejects a draft token at position $j$, we retain the remaining suffix of the draft $(dt_{j+1}, \dots, dt_m)$, which is used as the subsequent draft for future verification. After rejection, \ourmodel{} temporarily falls back to autoregressive decoding by the target LLM. Let the newly generated token sequence be $(t_1, \dots, t_k)$. We then perform \textit{span matching} between this generated sequence and the remaining draft sequence.

Formally, given the remaining draft $ (dt_{j+1}, \dots, dt_m)$, we search for a matching position $s$ such that:
\begin{equation}
\exists \, s \in [j+1, m-k+1], \quad (t_1, \dots, t_x) = (dt_s, \dots, dt_{s+x-1}).
\end{equation}
If such a position $s$ exists, we realign the draft sequence and resume parallel verification using the remaining suffix $(dt_{s+x}, \dots, dt_m)$. Otherwise, the model continues with standard autoregressive decoding.
This mechanism allows the LLM to recover alignment with the draft even after local mismatches, enabling efficient reuse of partially consistent draft tokens and further improving decoding efficiency. 
\section{Experiment Setups}
\subsection{Benchmarks}
We evaluate \ourmodel{} on two widely adopted repository-level code generation benchmarks.

\begin{itemize}
    \item \textbf{CoderEval}~\cite{codereval} contains 230 Python and Java tasks curated from real-world projects. It covers functions with diverse contextual dependencies and employs an automated execution platform to assess functional correctness. In our experiments, we use the Python subset, comprising 230 instances.
     
    \item \textbf{DevEval}~\cite{deveval} is a large-scale, manually annotated benchmark for repository-level code generation. It includes 1,825 test samples from 117 Python repositories spanning 10 popular domains (e.g., Internet and databases). However, as it does not provide a fully reproducible test environment, we configure the environment locally and obtain 1,462 valid data points by filtering instances based on whether the ground truth code successfully passes all associated tests.
\end{itemize}
\subsection{Baselines}
To evaluate the effectiveness of \ourmodel{}, we compare it against the following representative baselines:
% , including standard retrieval-augmented generation (RAG) with BM25, RL-Coder, RepoFormer, and Longcodezip.

\begin{itemize}
    \item \textbf{Standard RAG.} This baseline adopts a conventional BM25-based retrieval strategy. We use the target function signature along with its natural language description (e.g., comments) as the query, and retrieve relevant code contexts from the repository.

    \item \textbf{RL-Coder~\cite{rl-coder}.} This is an advanced repository-level code generation method that leverages reinforcement learning to optimize the retriever. It uses perplexity-based rewards to encourage the selection of informative code contexts.

    \item \textbf{RepoFormer~\cite{repoformer}.} It is a selective retrieval approach that first determines whether retrieval is necessary before generation. For a fair comparison, we adopt the StarCoder-1B that they have trained, which has a comparable parameter scale to the models used in our framework.

    \item \textbf{LongCodeZip~\cite{longcodezip}.} This is a code context compression and filtering method based on perplexity estimation. It employs a smaller model to compute perplexity scores for context selection. To ensure fairness, the same model used as the draft generator in \ourmodel{} is employed for perplexity computation in this baseline.
\end{itemize}

\subsection{Metrics}
\begin{itemize}
    \item \textbf{Pass@K.} We adopt the Pass@K metric to evaluate code generation performance, which measures the probability that at least one correct solution is produced within $K$ attempts:
    \begin{equation}
    \text{Pass@K} = 1 - \frac{\binom{n - c}{K}}{\binom{n}{K}},
    \end{equation}
    where $n$ is the total number of generated samples, $c$ is the number of correct solutions, and $K$ is the number of allowed attempts. Unless otherwise specified, we report Pass@1.
    \item \textbf{Time(s).} We measure generation efficiency in seconds using three metrics: (1) Preprocessing Time (Proc.T), which denotes the time required for context preprocessing after retrieval; (2) Generation Time (Gen.T), which measures the time taken by the LLM to generate code; and (3) Total Time (All.T), defined as the sum of Proc.T and Gen.T. 

    \item \textbf{Token.} We compute the length of the retrieved (or compressed) code context using the GPT tokenizer across all methods. This metric reflects the efficiency of code context selection and generation cost.
\end{itemize}

\subsection{Implementation Details}

\textbf{Backbone Models:} Due to the need to obtain the probability distribution from the LLM and constraints on computational resources, we select two open-source and widely used code LLM families: \textit{Qwen2.5-Coder} and \textit{DeepSeek-Coder}.
For \textit{Qwen2.5-Coder}, we adopt Qwen2.5-Coder-7B-Instruct as the generative (target) model and Qwen2.5-Coder-1.5B-Instruct as the draft model.
For \textit{DeepSeek-Coder}, we use DeepSeek-Coder-6.7B-Instruct as the generative model and DeepSeek-Coder-1.3B-Instruct as the draft model. For baseline methods, RAG and RL-Coder rely solely on the generative model without drafting models.
For RepoFormer, we replace the drafting model with its trained lightweight model of comparable scale, RepoFormer-1B.
For LongCodeZip and \ourmodel{}, we adopt the same generative--draft model combinations described above.

\textbf{Hardware and Implementation Details:}
All experiments are conducted on a server equipped with 4 NVIDIA RTX 4090 GPUs, which are used for both draft and generative LLM inference.
The CPU is an Intel\textsuperscript{\textregistered} Xeon\textsuperscript{\textregistered} Gold 6330. All time tests are conducted under the above conditions to ensure a fair comparison. For \ourmodel{}, based on our preliminary experiments we set the hyperparameter $L$ in the Logic Similarity Perspective is 1 and the size of each group in the \textit{Map Phrase} is 4. All other baselines follow the default configurations reported in their default settings. 
To ensure deterministic generation and eliminate randomness, the decoding temperature is set to 0.
% The decoding temperature is 0. 
To verify the effectiveness of different numbers of retrieval blocks, the number of retrieved contexts is varied among $\{0, 3, 5, 7, 10\}$, and the corresponding results are reported. The selection of these values is guided by preliminary experiments, which reveal a clear turning point within this range: the Pass@1 score first improves with increasing $k$ and then degrades as more contexts are introduced. This trend aligns with our motivation, indicating that while additional context can provide useful information, excessive retrieval introduces noise that negatively affects generation quality.

\section{Experimental Results}
We aim to address the following three research questions:
\begin{itemize}
    \item \textbf{RQ1: Code Generation Quality.} How does \ourmodel{} perform in terms of the correctness of generated code for repository-level code generation tasks?
    % How does \ourmodel{} contribute to the correctness of generated repository-level code?
    \item \textbf{RQ2: Efficiency and Cost Analysis.} How does \ourmodel{} perform with respect to computational cost (e.g., number of input tokens) and efficiency (e.g., generation time)?
    % How does \ourmodel{} affect the cost (number of input tokens) and efficiency (generation time)?
    \item \textbf{RQ3: Ablation Study.} How does each component of \ourmodel{} contribute to the overall performance?
\end{itemize}

\subsection{RQ1: Code Generation Quality}
% \begin{table}[t]
% \centering
% \small
% \resizebox{0.9\linewidth}{!}{
% \begin{tabular}{llccccc ccccc}
% \toprule
% \multirow{2}{*}{Model} & \multirow{2}{*}{Method} & \multicolumn{5}{c}{CoderEval} & \multicolumn{5}{c}{RepoExec} \\
% \cmidrule(lr){3-7} \cmidrule(lr){8-12}
%  &  & k=0 & k=5 & k=10 & k=15 & k=20 & k=0 & k=5 & k=10 & k=15 & k=20 \\
% \midrule

% \multirow{5}{*}{Qwen2.5-Coder} 
%  & Full-Context & 24.4  & 37.0  & 33.5  & 33.0  & 33.5   &  \\
%  & RL-Coder     & 24.4  &         &         &         &          &  \\
%  & Repoformer   & 24.4  & 30.9  & 29.6  & 29.6  & 28.3   &  \\
%  & LongCodeZip  & 24.4  &         &         &         &          &  \\
%  & \ourmodel{}  & 24.4  & \textbf{40.4} &         &         &          &  \\

% \midrule

% \multirow{5}{*}{DeepSeek-Coder} 
%  & Full-Context & 24.4 &  &  &  &  &  \\
%  & RL-Coder & 24.4  &  &  &  &  &  \\
%  & Repoformer & 24.4 &  &  &  &  &  \\
%  & LongCodeZip & 24.4 &  &  &  &  &  \\
%  & \ourmodel{} & 24.4 &  &  &  &  &  \\

% \bottomrule
% \end{tabular}
% }
% \caption{Performance comparison on CoderEval and RepoExec.}
% \label{tab:main_results}
% \end{table}

\begin{table}[t]
\centering
\setlength{\abovecaptionskip}{0.1cm}
\caption{Pass@1 results comparison on CoderEval and DevEval. The green percentage boxes indicate the percentage improvement in Pass@1 for \ourmodel{} compared to the RAG baseline. The text in bold indicates the highest score, and the underlined values denote the performance of the best baseline.}
\resizebox{\linewidth}{!}{
\begin{tabular}{llccccc ccccc}
\toprule
\multirow{2}{*}{Model} & \multirow{2}{*}{Method} & \multicolumn{5}{c}{CoderEval} & \multicolumn{5}{c}{DevEval} \\
\cmidrule(lr){3-7} \cmidrule(lr){8-12}
 &  & K=0 & K=3 & K=5 & K=7 & K=10 & K=0 & K=3 & K=5 & K=7 & K=10 \\
\midrule

\multirow{5}{*}{Qwen2.5-Coder} 
 & RAG & 24.4  & {31.7}  & {37.0}  & {35.2}  & {33.5}   & 13.0 & {15.8} & \uline{15.9} & {15.0} &  {15.3}\\
 & RL-Coder     & 24.4  & 26.5 &  30.0  & 27.0  & 31.3  & 13.0 & \uline{16.2} & \uline{15.9} & 15.7 & 16.1\\
 & Repoformer   & 24.4  & 29.6  & 30.9  & 31.3  & 29.6   & 13.0& 13.4 &13.0 &12.7 &13.0 \\
 & LongCodeZip  & 24.4  & \uline{36.5}  & \uline{38.3}  & \uline{36.5}  & \uline{35.2}   &13.0& 15.3 & 15.8 & \uline{16.1} &\uline{16.3} \\
 & \ourmodel{}  & 24.4  & \textbf{37.8\ua{19.2\%}}  &\textbf{40.0\ua{8.1\%}} &  \textbf{39.6\ua{12.5\%}} &\textbf{38.7\ua{15.5\%}} & 13.0 & \textbf{19.9\ua{25.9\%}} &  \textbf{19.0\ua{19.5\%}} & \textbf{19.3\ua{28.7\%}} & \textbf{18.1\ua{18.3\%}} \\

\midrule

\multirow{5}{*}{DeepSeek-Coder} 
 & RAG & 25.2    & \uline{31.3} & \uline{30.0}   & {28.3}   & {23.9} &  13.4 &  {24.1} &  {23.2} & {24.5} & {23.5}\\
 & RL-Coder     & 25.2    & 26.5   &  27.4    & 27.8    & 26.3   & 13.4 &\uline{\textbf{24.4}} & \uline{\textbf{26.1}}  &   24.7  & 24.2\\
 & Repoformer   & 25.2    & 28.7 & 29.6    & \uline{29.1}   & 27.4 &  13.4 & 18.0 & 17.7& 17.9&18.1 \\
 & LongCodeZip  & 25.2    & 26.5 & \uline{30.0}    & 26.9   & \uline{27.8} &  13.4 & 22.9 & 24.4 & \uline{\textbf{27.6}} &  \uline{\textbf{27.6}}\\
 & \ourmodel{}  & 25.2    & \textbf{31.7\ua{1.3\%}} & \textbf{35.2\ua{17.3\%}}    & \textbf{32.6\ua{15.2\%}}  & \textbf{36.5\ua{52.7\%}} & 13.4 & 24.1\ua{0.0\%} & 25.5\ua{10.0\%} & 26.3\ua{7.3\%} & 26.3\ua{11.9\%}\\
\bottomrule
\end{tabular}
}
\label{tab:main_results}
\end{table}

Table \ref{tab:main_results} presents Pass@1 scores of the generated code for both the baseline methods and \ourmodel{} on DevEval and CoderEval. Except for RL-Coder, which employs a separately trained retriever, all methods adopt BM25 for retrieval.

Standard RAG baselines do not exhibit a monotonic improvement in Pass@1 as the number of retrieved context blocks increases. For both Qwen2.5-Coder and DeepSeek-Coder, performance typically shows an initial improvement followed by degradation, or fluctuates across different values of $K$. For example, Qwen2.5-Coder on CoderEval improves from $K=3$ to $K=5$, but declines at $K=7$, while DeepSeek-Coder achieves its best performance at $K=7$ on DevEval. These observations indicate that although increasing $K$ introduces more potentially useful context, it also brings in noisy or irrelevant information, which ultimately hinders the model’s ability to effectively utilize the retrieved context.

Existing methods attempt to mitigate this issue from different perspectives, yet their effectiveness remains limited. Specifically, RL-Coder improves retrieval quality via reinforcement learning, but its gains are marginal—for instance, on DevEval it achieves less than a 1\% improvement over standard RAG for Qwen2.5-Coder. Moreover, its reliance on specific training dataset leads to limited generalization, resulting in inferior performance compared to RAG on CoderEval. RepoFormer takes a different approach by learning to decide whether external context is needed through special control tokens; however, its performance still falls behind RAG, suggesting that the key challenge is not simply deciding whether to retrieve context, but rather how to filter and utilize it effectively. LongCodeZip further explores fine-grained selection by using a small LLM to estimate perplexity and select informative code lines. While it performs competitively at larger $K$, its reliance on perplexity-based estimation is inherently imprecise and lacks interpretability.

In contrast, \ourmodel{} addresses these limitations by explicitly modeling context relevance through the SADGS algorithm, which integrates two complementary signals: API call relationships and structural similarity between draft generations and repository code. This design enables \ourmodel{} to provide concise and high-quality context, thereby reducing interference from irrelevant information and improving the model’s ability to leverage useful context during generation.
As a result, \ourmodel{} achieves strong and stable performance across benchmarks. On CoderEval, both LLM combinations attain the best results among all baselines, with DeepSeek-Coder improving Pass@1 by nearly 52.7\% over standard RAG at $K=10$ and 31.2\% over best baseline LongCodeZip. On DevEval, \ourmodel{} consistently outperforms the RAG baseline across all settings. Although it is slightly inferior to RL-Coder ($K=3, 5$) and LongCodeZip ($K=7, 10$) at specific values of $K$, \ourmodel{} maintains significantly more stable performance as $K$ varies. These results demonstrate that \ourmodel{} not only improves code generation quality but also provides a more interpretable and reliable mechanism for context selection. Moreover, \ourmodel{} consistently outperforms standard RAG across different $K$ values and exhibits a similar performance trend (i.e., first improving and then degrading as $K$ increases). This suggests that the optimal $K$ identified under RAG can be directly applied to \ourmodel{} to achieve the best performance.
\begin{SummaryBox}
\textbf{RQ1 Summary: } \ourmodel{} consistently achieves strong and stable performance in terms of code correctness across both CoderEval and DevEval, outperforming or matching competitive baselines. By modeling context relevance via code structure and API relationships, \ourmodel{} filters noisy retrievals and provides concise, high-quality context, improving generation and interpretability. These results demonstrate that precise, structured context selection is critical for repository-level code generation.
% By explicitly modeling context relevance through code structure and API call relationships, \ourmodel{} effectively filters out noisy or irrelevant retrievals and provides concise, high-quality context. This not only improves the model’s ability to utilize retrieved information for code generation, but also enhances interpretability. Overall, the results demonstrate that precise and structured context selection is critical for repository-level code generation, and \ourmodel{} offers a more effective solution to this challenge.
\end{SummaryBox}
% \ourmodel{} achieves competitive performance in terms of time efficiency and token cost while maintaining optimal quality, even outperforming Full-Context. CtxMem substantially reduces both time and token cost, whereas SessAST introduces additional cost and inference time to handle fogetting.
\subsection{RQ2: Efficiency and Cost Analysis}
\begin{figure*}
	\centering
    \setlength{\abovecaptionskip}{0.1cm}
	\includegraphics[width=1\textwidth]{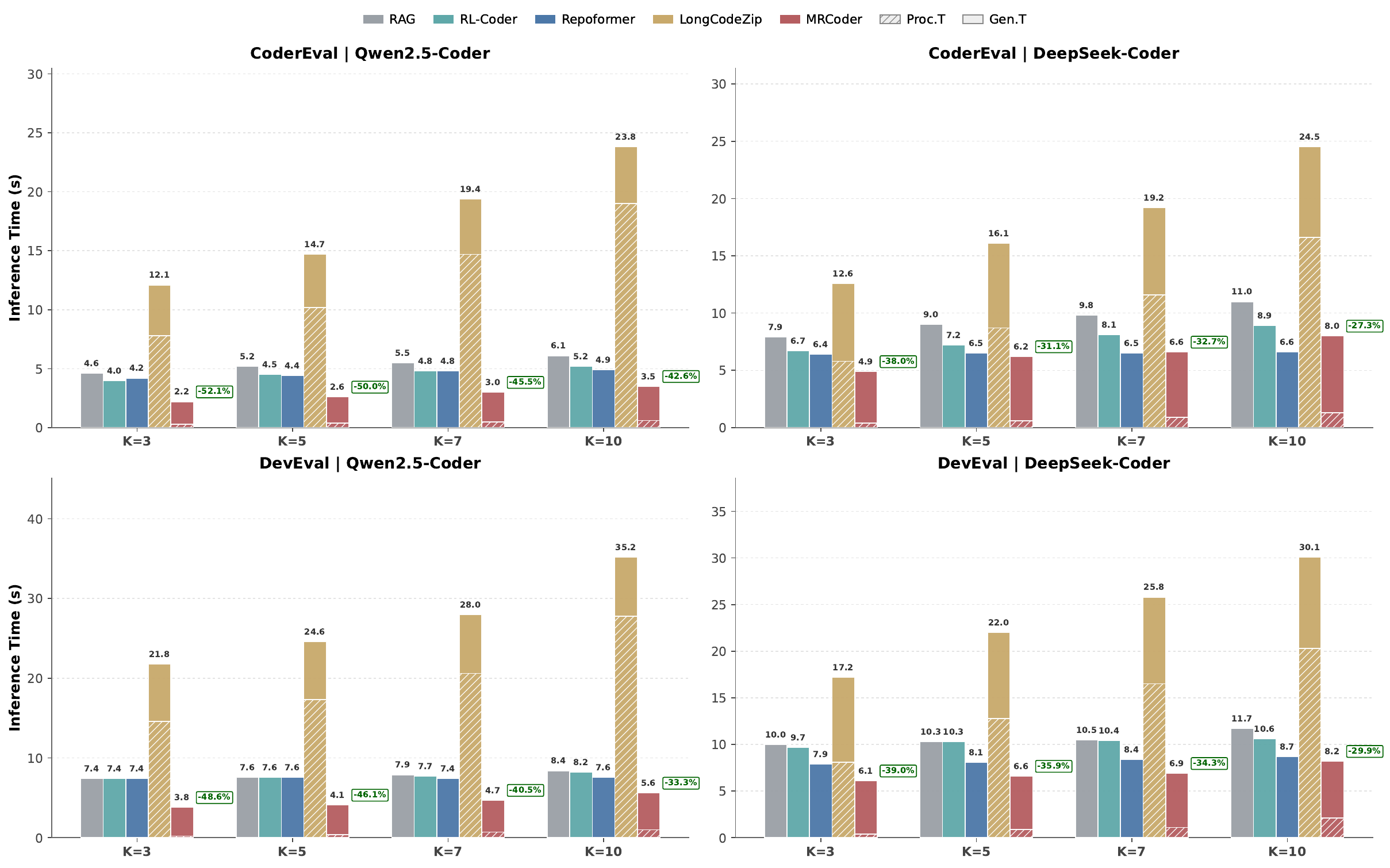}
 	\caption{Inference time comparison of \ourmodel{} and baseline methods. The green percentage boxes indicate the percentage reduction in time for \ourmodel{} compared to the RAG baseline. 
    % The text in bold indicates the lowest time usage, and the underlined values denote the performance of the RAG baseline.
    }
	\label{fig:rq2_time}
    \vspace{-0.3cm}
\end{figure*}

\begin{table}[t]
\centering
\setlength{\abovecaptionskip}{0.1cm}
\caption{Token consumption comparison on CoderEval and DevEval. The green percentage boxes indicate the percentage reduction in tokens for \ourmodel{} compared to the RAG baseline. The text in bold indicates the lowest token usage overall, and the underlined values denote the best-performing baseline in terms of token efficiency.}
\resizebox{\linewidth}{!}{
\begin{tabular}{llcccc cccc}
\toprule
\multirow{2}{*}{Model} & \multirow{2}{*}{Method} & \multicolumn{4}{c}{CoderEval} & \multicolumn{4}{c}{DevEval} \\
\cmidrule(lr){3-6} \cmidrule(lr){7-10}
 &  & K=3 & K=5 & K=7 & K=10 & K=3 & K=5 & K=7 & K=10 \\
\midrule

\multirow{5}{*}{Qwen2.5-Coder} 
 & RAG    & 1.7k & 3.1k & 4.6k & 7.3k & 1.8k & 3.2k & 4.6k & 7.0k \\
 & RL-Coder      & 1.0k & 2.0k & 3.3k & 5.6k & 1.3k & 2.4k & 3.6k & 5.5k \\
 & Repoformer      & \uline{\textbf{0.7k}} & \uline{\textbf{1.3k}} & \uline{\textbf{1.9k}} & \uline{\textbf{2.8k}} & \uline{\textbf{0.8k}} & \uline{\textbf{1.3k}} & 1.9k & 3.0k \\
 & LongCodeZip    & 0.8k & \uline{\textbf{1.3k}} & \uline{\textbf{1.9k}} & 2.9k & 0.9k & \uline{\textbf{1.3k}} & \uline{\textbf{1.8k}} & \uline{\textbf{2.6k}} \\
 & \ourmodel{}   & 1.2k\df{29.4\%} & 1.9k\df{38.7\%} & 2.7k\df{41.3\%} & 3.6k\df{50.7\%} & 1.2k\df{33.3\%} & 1.8k\df{43.8\%} & 2.7k\df{41.3\%} & 3.8k\df{45.7\%} \\

\midrule

\multirow{5}{*}{DeepSeek-Coder} 
 & RAG     & 1.7k & 3.1k & 4.6k & 7.3k & 1.8k & 3.2k & 4.6k & 7.0k \\
 & RL-Coder   & 1.0k & 2.0k & 3.3k & 5.6k & 1.3k & 2.4k & 3.6k & 5.5k \\
 & Repoformer     & \uline{\textbf{0.7k}} & 1.3k & 1.9k & 2.8k & \uline{\textbf{0.8k}} & \uline{\textbf{1.3k}} & 1.9k & 3.0k \\
 & LongCodeZip    & \uline{\textbf{0.7k}} & \uline{\textbf{1.2k}} & \uline{\textbf{1.7k}} & \uline{\textbf{2.5k}} & 0.9k & \uline{\textbf{1.3k}} & \uline{\textbf{1.8k}} & \uline{\textbf{2.5k}} \\
 &  \ourmodel{}& 1.0k\df{41.2\%} & 1.8k\df{41.9\%} & 2.2k\df{52.2\%} & 3.5k\df{52.1\%} & 1.1k\df{38.9\%} & 2.0k\df{37.5\%} & 2.5k\df{45.6\%} & 3.4k\df{51.4\%} \\

\bottomrule
\end{tabular}
}
\label{tab:rq2_token}
\end{table}

% Tables \ref{tab:rq2_time_codereval} and \ref{tab:rq2_time_deveval} present the inference time comparisons across different methods on CoderEval and DevEval.
Figure \ref{fig:rq2_time} presents the inference time comparisons and Table \ref{tab:rq2_token} presents the token comparisons across different methods on CoderEval and DevEval. Standard RAG exhibits a clear increase in both inference time and token consumption as the number of retrieval blocks $K$ grows. Since it directly concatenates all retrieved contexts, larger $K$ leads to longer input sequences, which significantly increases generation latency and computational cost. This trend is consistent across both Qwen2.5-Coder and DeepSeek-Coder, indicating that naive retrieval scaling introduces substantial inefficiency.

Existing baselines attempt to reduce cost from different perspectives, but each has inherent limitations. RL-Coder reduces token consumption compared to RAG by improving retrieval quality, which slightly lowers generation time. We attribute this to the fact that, although RL-Coder does not directly train the retriever to compress context, it assigns higher scores to function-level code contexts that are more semantically consistent with the current query, while assigning lower retrieval scores to potential longer external API contexts (such as classes), thereby reducing token costs and generation time. However, its improvements remain moderate, as it still relies on incorporating multiple retrieved contexts. RepoFormer achieves the lowest token consumption by aggressively controlling whether retrieval is used (only one token's inference time is required), resulting in minimal processing overhead and stable inference time. Nevertheless, this strategy sacrifices performance, as shown in RQ1, indicating it cannot achieve a trade-off between inference cost and time versus quality. LongCodeZip adopts a fine-grained selection strategy and achieves competitive token reduction at larger $K$, but its computational overhead is substantial due to multiple forward passes of a small model for perplexity estimation. As a result, its processing time increases dramatically with $K$, leading to the highest overall latency among all methods. This limits its applicability in time-sensitive code generation scenarios.

In contrast, \ourmodel{} achieves a more favorable balance between efficiency and effectiveness. Although \ourmodel{} introduces additional overhead—stemming from draft generation by a small model in the \textit{Map Phase} and context selection via SADGS—the batching and parallelization design ensures that the overall latency increase remains limited. As shown in Figure \ref{fig:rq2_time}, it incurs at most around 2 seconds of additional inference time. Moreover, the use of parallel decoding for draft verification in the \textit{Reduce Phase} significantly improves inference efficiency, effectively offsetting the extra computation introduced in earlier stages. On CoderEval, \ourmodel{} reduces total inference time by up to 52.1\% for Qwen2.5-Coder and 38.0\% for DeepSeek-Coder at $k=3$, while maintaining consistent reductions as $K$ increases. Similar trends are observed on DevEval, where time reductions reach up to 48.6\% and 39.0\%, respectively.
In most experimental settings, \ourmodel{} achieved the highest generation efficiency compared to all baseline methods.
Notably, the processing overhead introduced by \ourmodel{} remains minimal compared to LongCodeZip, demonstrating that its selection mechanism is lightweight and scalable. In terms of token efficiency, \ourmodel{} consistently reduces token consumption by 30\%–50\% compared to RAG across both benchmarks and model settings. While it does not always achieve the absolute lowest token usage (as RepoFormer and LongCodeZip are more aggressive in pruning), it maintains substantially lower token cost without sacrificing generation quality. This highlights that \ourmodel{} avoids the inefficiency of redundant context while preserving the essential information required for accurate code generation.

% Overall, these results demonstrate that \ourmodel{} effectively mitigates the cost explosion caused by increasing $K$, achieving significant reductions in both inference time and token consumption. Compared to prior methods, it provides a more balanced and scalable solution, improving efficiency while maintaining high generation quality.

\begin{SummaryBox}
\textbf{RQ2 Summary: } \ourmodel{} improves both efficiency and generation quality simultaneously. By effectively filtering redundant context, it reduces token consumption by 30\%–50\% and inference time by up to over 50\%, mitigating the cost explosion of increasing $K$. Compared to prior methods, it avoids heavy computational overhead and achieves the highest generation efficiency, providing a more efficient and scalable solution for repository-level code generation.
\end{SummaryBox}
\subsection{RQ3: Ablation Study}

\begin{table}[t]
\centering
\small
\setlength{\abovecaptionskip}{0.1cm}
\caption{Pass@1 ablation study results on CoderEval and DevEval. Red and green boxes represent the absolute decrease and increase, respectively, when specific components are removed from \ourmodel{}.}
\begin{tabular}{llcccccccc}
\toprule
\multirow{2}{*}{Model} & \multirow{2}{*}{Method} & \multicolumn{4}{c}{CoderEval} & \multicolumn{4}{c}{DevEval} \\
\cmidrule(lr){3-6} \cmidrule(lr){7-10}
 & & K=3 & K=5 & K=7 & K=10 & K=3 & K=5 & K=7 & K=10 \\
\midrule
\multirow{3}{*}{Qwen2.5-Coder} 
 & \ourmodel{} & 37.8 & 40.0 & 39.6 & 38.7 & 19.9 & 19.0 & 19.3 & 18.1 \\
 & w/o\_API & 35.2\da{2.6} & 39.6\da{0.4} & 39.6\da{0.0} & 37.0\da{1.7} & 14.8\da{5.1} & 14.7\da{4.3} & 16.0\da{3.3} & 14.8\da{3.3} \\
 & w/o\_Sim & 36.5\da{1.3} & 38.3\da{1.7} & 39.6\da{0.0} & 37.8\da{0.9} & 17.3\da{2.6} & 17.2\da{1.8} & 17.1\da{2.2} & 16.9\da{1.2} \\
\midrule
\multirow{3}{*}{DeepSeek-Coder} 
 & \ourmodel{} & 31.7 & 35.2 & 32.6 & 36.5 & 24.1 & 25.5 & 26.3 & 26.3 \\
 & w/o\_API     & 32.2\ua{0.5} & 36.5\ua{1.3} & 31.7\da{0.9} & 36.0\da{0.5} & 20.9\da{3.2} & 23.1\da{2.4} & 23.1\da{3.2} & 24.6\da{1.7} \\
 & w/o\_Sim     & 31.3\da{0.4} & 34.8\da{0.4} & 33.5\ua{0.9} & 33.9\da{2.6} & 22.4\da{1.7} & 24.1\da{1.4} & 24.1\da{2.2} & 25.2\da{1.1} \\
\bottomrule
\end{tabular}
\label{tab:rq3_ablation_pass@1}
\end{table}

\begin{table}[t]
\centering
\setlength{\abovecaptionskip}{0.1cm}
\small
\caption{Token ablation consumption of \ourmodel{} on CoderEval and DevEval. The green boxes indicate the reduction in context length when removing specific modules.}
\begin{tabular}{llcccccccc}
\toprule
\multirow{2}{*}{Model} & \multirow{2}{*}{Method} & \multicolumn{4}{c}{CoderEval} & \multicolumn{4}{c}{DevEval} \\
\cmidrule(lr){3-6} \cmidrule(lr){7-10}
 & & K=3 & K=5 & K=7 & K=10 & K=3 & K=5 & K=7 & K=10 \\
\midrule
\multirow{3}{*}{Qwen2.5-Coder} 
 & \ourmodel{} & {1.2k} & {1.9k} & {2.7k} & {3.6k} & {1.2k }&{1.8k} & {2.7k} &{ 3.8k}  \\
 & w/o\_API     & 0.6k\df{50.0\%} & 1.3k\df{31.6\%} & 1.4k\df{48.1\%} & 2.1k\df{41.7\%} & 0.8k\df{33.3\%} & 1.6k\df{11.1\%} & 1.9k\df{29.6\%} & 2.5k\df{34.2\%} \\
 & w/o\_Sim     & 1.0k\df{16.7\%} & 1.6k\df{15.8\%} & 2.3k\df{14.8\%} & 3.0k\df{16.7\%} & 0.8k\df{33.3\%} & 1.7k\df{5.6\%} & 1.3k\df{51.9\%} & 2.3k\df{39.5\%} \\
\midrule
\multirow{3}{*}{DeepSeek-Coder} 
 & \ourmodel{} & {1.0k} & {1.8k} & {2.2k} & {3.5k} & {1.1k}  & {2.0k} &  {2.5k} & {3.4k}\\
 & w/o\_API     & 0.5k\df{50.0\%} & 1.3k\df{27.8\%} & 1.1k\df{50.0\%} & 2.0k\df{42.9\%} & 0.9k\df{18.2\%} & 1.6k\df{20.0\%} & 1.8k\df{28.0\%} & 2.4k\df{29.4\%} \\
 & w/o\_Sim     & 0.9k\df{10.0\%} & 1.5k\df{16.7\%} & 1.9k\df{13.6\%} & 3.0k\df{14.3\%} & 0.8k\df{27.3\%} & 1.2k\df{40.0\%} & 1.6k\df{36.0\%} & 2.2k\df{35.3\%} \\
\bottomrule
\end{tabular}
\label{tab:rq3_ablation_token}
\vspace{-0.3cm}
\end{table}

% \begin{table}[t]
% \centering
% \small
% \caption{Token comsumption comparison on CoderEval and DevEval.}
% \begin{tabular}{llcccccccc}
% \toprule
% \multirow{2}{*}{Model} & \multirow{2}{*}{Method} & \multicolumn{4}{c}{CoderEval} & \multicolumn{4}{c}{DevEval} \\
% \cmidrule(lr){3-6} \cmidrule(lr){7-10}
%  & & k=3 & k=5 & k=7 & k=10 & k=3 & k=5 & k=7 & k=10 \\
% \midrule
% \multirow{3}{*}{Qwen2.5-Coder} 
%  & \ourmodel{} & 1.2k & 1.9k & 2.7k & 3.6k & 1.2k & 1.8k & 2.7k & 3.8k  \\
%  & w/o\_API     & 0.6k & 1.3k & 1.4k & 2.1k & 0.8k & 1.6k & 1.9k & 2.5k \\
%  & w/o\_Sim     & 1.0k & 1.6k & 2.3k & 3.0k & 0.8k & 1.7k & 1.3k & 2.3k \\
% \midrule
% \multirow{3}{*}{DeepSeek-Coder} 
%  & \ourmodel{} & 1.0k & 1.8k & 2.2k & 3.5k & 1.1k  & 2.0k &  2.5k & 3.4k\\
%  & w/o\_API     & 0.5k & 1.3k & 1.1k & 2.0k & 0.9k & 1.6k& 1.8k & 2.4k \\
%  & w/o\_Sim     & 0.9k & 1.5k & 1.9k & 3.0k & 0.8k & 1.2k & 1.6k & 2.2k \\
% \bottomrule
% \end{tabular}
% \end{table}
\begin{figure*}
	\centering
    \setlength{\abovecaptionskip}{0.1cm}
	\includegraphics[width=1\textwidth]{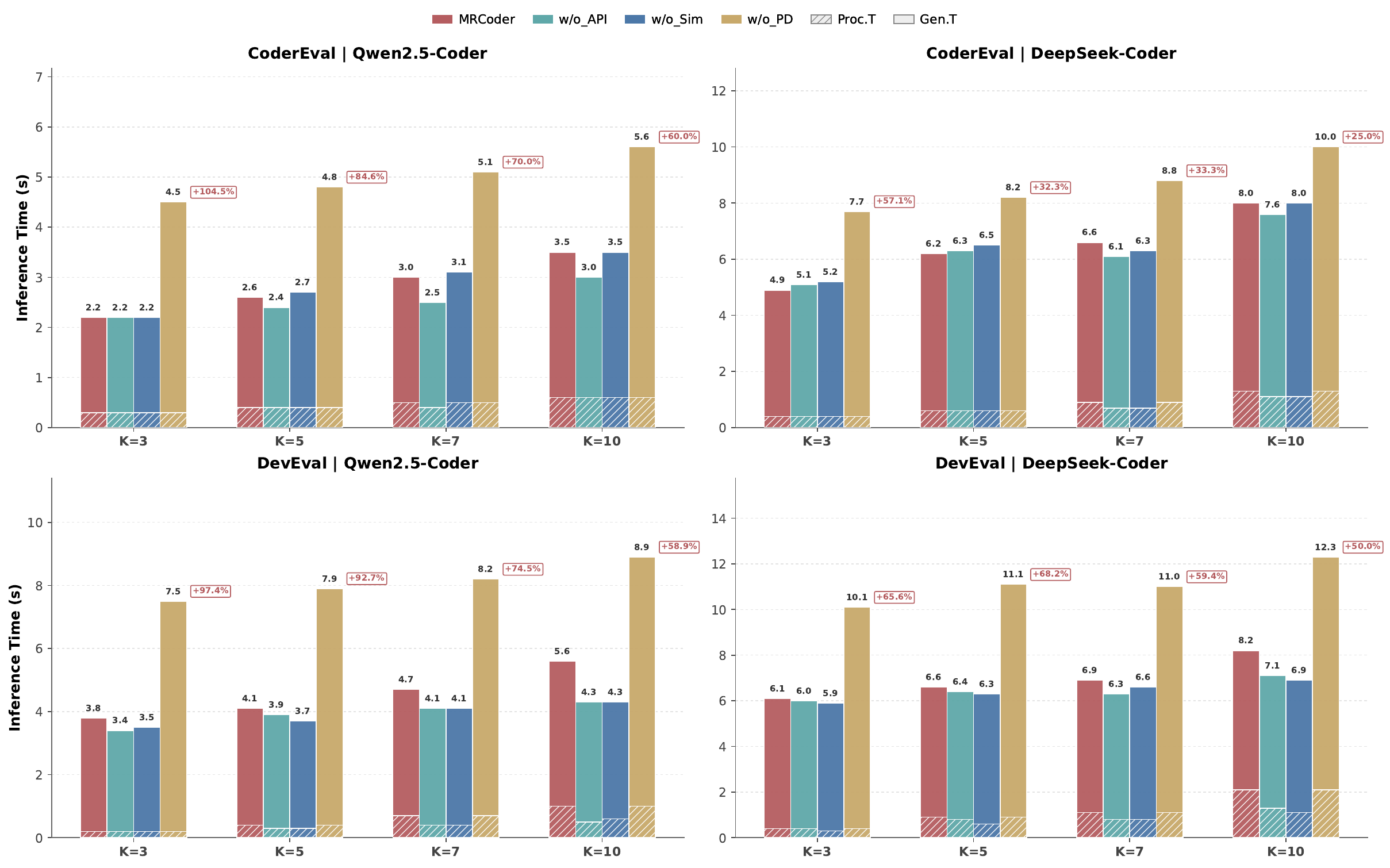}
 	\caption{
    % Inference time performance across different experimental settings on CoderEval and DevEval. The green percentage boxes indicate the \textcolor{red}{percentage reduction in time for \ourmodel{} compared to the RAG baseline}. 
    Efficiency ablation analysis for \ourmodel{} on CoderEval and DevEval. The red percentages boxes in w/o\_PD denote the percentage reduction in time compared to \ourmodel{}.
    }
	\label{fig:rq3_time}
    \vspace{-0.3cm}
\end{figure*}
To answer this question, we conduct ablation studies to analyze the contribution of each component in \ourmodel{}.
Specifically, we remove individual components, including API-based selection, similarity-based selection, and parallel decoding (PD), to evaluate their impact on overall performance in terms of both code quality and efficiency/cost. Since parallel decoding does not affect code quality, the quality-focused analysis considers only the removal of the two selection components.
% including API-based selection, similarity-based selection, and parallel decoding (PD). 

From the perspective of code quality, as shown in Table \ref{tab:rq3_ablation_pass@1}, removing either the API-based or similarity-based selection leads to consistent performance degradation across most settings. Specifically, removing the API component results in more significant drops on DevEval (e.g., up to 5.1 absolute decrease for Qwen2.5-Coder at $K=3$), indicating that API call relationships play a crucial role in identifying functionally relevant context. In contrast, removing the similarity component leads to relatively smaller but still consistent declines, suggesting that structural similarity provides complementary signals that refine context selection. Although occasional marginal improvements are observed in certain settings (e.g., DeepSeek-Coder at specific $K$ values), these gains are not stable. 
% This phenomenon is particularly evident in DeepSeek-Coder’s performance. 
% \textcolor{red}{We attribute this to its relatively weaker ability to utilize retrieved context: even after context filtering by \ourmodel{}, a small amount of residual noise may remain, to which DeepSeek-Coder is more sensitive.} 
As a result, slight improvements can be observed in both w/o\_API and w/o\_Sim variants in some cases. However, the overall performance trends consistently demonstrate that both components contribute synergistically to improving code generation quality.

In terms of inference time, as shown in Figure \ref{fig:rq3_time}, the differences between \ourmodel{} and its w/o\_API and w/o\_Sim variants are relatively minor, suggesting that both selection mechanisms introduce negligible computational overhead and remain lightweight. For the perspective of cost, as shown in Table \ref{tab:rq3_ablation_token} , the impact of each component exhibits different characteristics. Removing the API or similarity modules consistently reduces token consumption, as fewer constraints are applied during context selection, leading to shorter input sequences. However, this reduction comes at the cost of degraded generation quality, indicating that aggressive pruning without structural guidance harms effectiveness.
In contrast, removing parallel decoding (w/o\_PD) results in a substantial increase in inference latency across all settings. For example, total inference time increases by up to 104.5\% on CoderEval and nearly doubles on DevEval. This demonstrates that parallel decoding is the key factor enabling efficient generation in \ourmodel{}, not only effectively amortizing the additional overhead introduced in earlier stages such as draft generation and context selection, but improving the overall generation efficiency.

% Overall, these results highlight that API-based and similarity-based selection are essential for maintaining high generation quality, while parallel decoding is critical for ensuring efficiency. The combination of these components allows \ourmodel{} to achieve a balanced trade-off between effectiveness and efficiency.
\begin{SummaryBox}
\textbf{RQ3 Summary: } Both API-based and similarity-based selection are essential for achieving high-quality code generation, contributing complementary signals for effective context filtering. While removing these components may reduce token consumption, it leads to noticeable performance degradation, highlighting the importance of these two complementary components. In contrast, parallel decoding plays a critical role in efficiency, significantly reducing inference latency and offsetting the overhead of earlier stages. Together, these components enable \ourmodel{} to maintain strong generation quality while achieving efficient and scalable inference.
\end{SummaryBox}

\section{Disscussion}
\subsection{Case Study}
\begin{figure*}
	\centering
    \setlength{\abovecaptionskip}{0.1cm}
	\includegraphics[width=1\textwidth]{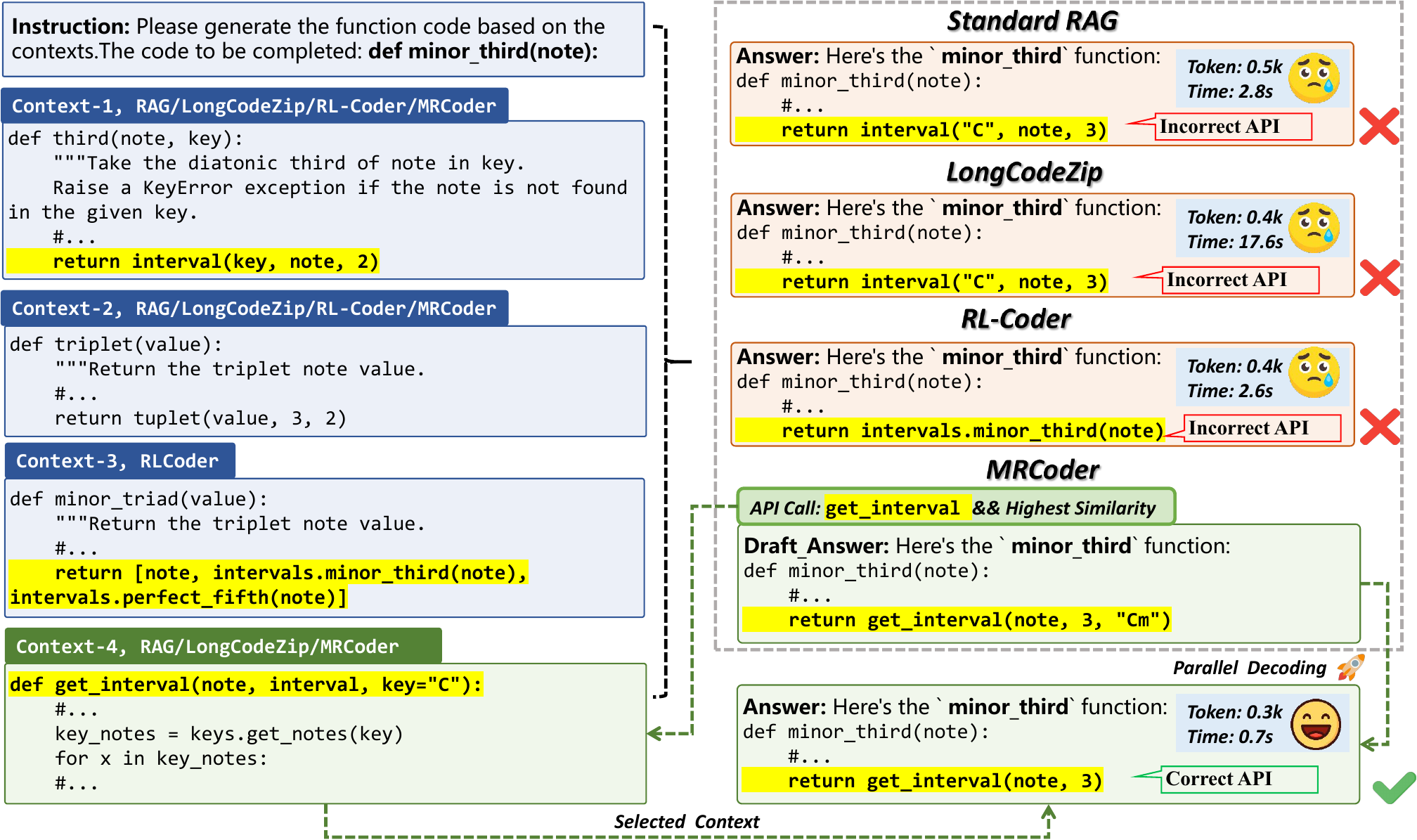}
 	\caption{An example from the DevEval benchmark, illustrating the code generation results and corresponding retrieved contexts for \ourmodel{} and the baselines (standard RAG, LongCodeZip, and RL-Coder), using Qwen2.5-Coder as the backbone model. 
    Standard RAG, LongCodeZip, and MRCoder (Map phase) employ BM25 to retrieve Contexts 1, 2, and 4, while RL-Coder retrieves Contexts 1, 2, and 3 using its retriever. For MRCoder, Context 4 is ultimately selected from the context group (Contexts 1, 2, 4) and is then used in the Reduce phase.
    % Standard RAG, LongCodeZip and MRCoder (Map Phrase) employ BM25 for retrieval, corresponding to Contexts 1, 2, and 4, while RL-Coder retrieves Contexts 1, 2, and 3 using its retriever. The Context 4 is used in Reduce Phrase in MRCoder.
    }
	\label{fig:case_study}
    \vspace{-0.3cm}
\end{figure*}

Figure \ref{fig:case_study} presents a representative example from the DevEval benchmark, comparing the code generation results of \ourmodel{} and the baselines.
% demonstrating the effectiveness of \ourmodel{} in selecting functionally relevant contexts for repository-level code generation. 
The task is to implement the function \textit{minor\_third(note)} based on multiple retrieved code snippets.
% with overlapping but potentially misleading semantics.

As shown in the figure, in Standard RAG and other baselines (LongCodeZip and RL-Coder), the LLM directly utilize all their retrieved contexts, leading to incorrect API usage. Specifically, they either invoke \textit{interval("C", note, 3)} or \textit{intervals.minor\_third(note)}, both of which are inconsistent with the repository’s actual API design. This error stems from the presence of superficially similar but semantically misaligned contexts (e.g., \textit{third},\textit{ triplet}, and \textit{minor\_triad}), which introduce ambiguity and misguide the generation process. In contrast, \ourmodel{} employs a draft-guided selection mechanism to identify the most relevant context. During the \textit{Map Phase}, the draft LLM generates an intermediate solution that correctly reflects the intended API pattern (\textit{get\_interval}). Based on API call consistency and logical similarity, the LLM successfully filters out noisy contexts and retains the key snippet defining \textit{def get\_interval(note, interval, key="C")}. In the \textit{Reduce Phase}, the refined context is used by the target model to generate the final implementation, yielding the correct API call. 

Beyond correctness, \ourmodel{} also achieves notable efficiency gains. As shown in the figure, baseline methods process longer input contexts (0.3k–0.5k tokens) and incur higher latency (2.6s–17.6s), particularly for methods such as LongCodeZip that require multiple forward passes. In contrast, \ourmodel{} reduces the input to only the most relevant context (0.3k tokens) and, combined with parallel decoding, significantly lowers the inference time to 0.7s. 

This case demonstrates that \ourmodel{} not only improves functional accuracy but also effectively reduces computational overhead by eliminating redundant context and enabling more efficient decoding.

\subsection{Impact of Context Partition Strategy in Map Phrase}
\label{diss:partition_strategy}
\begin{table}[t]
\centering
\small
\setlength{\abovecaptionskip}{0.1cm}
\caption{Pass@1 results comparison of different context partition strategies in the Map phase. The text in bold indicates the highest score.}
\begin{tabular}{llcccccccc}
\toprule
\multirow{2}{*}{Model} & \multirow{2}{*}{Partition Strategy} 
& \multicolumn{4}{c}{CoderEval} 
& \multicolumn{4}{c}{DevEval} \\
\cmidrule(lr){3-6} \cmidrule(lr){7-10}
& & K=3 & K=5 & K=7 & K=10 & K=3 & K=5 & K=7 & K=10 \\
\midrule

\multirow{4}{*}{Qwen2.5-Coder}
& Random Partition      & 36.1 & 38.7 & 38.3 & \textbf{42.2} & 19.9 & 19.0 & 19.6 & 20.0 \\
& Interleaved Partition & 36.4 & 38.3 & 38.3 & \textbf{43.4} & 19.6 & 19.5 & 19.6 & 20.1 \\
& Clustering Partition  & 36.3 & 39.6 & 38.3 & 37.8 &19.7& 19.7 & 19.9 & 18.9 \\
& Sequential Partition  & \textbf{37.8} & \textbf{40.0} & \textbf{39.6} & 38.7 & \textbf{19.9} & \textbf{19.0} & \textbf{19.3} & \textbf{18.1} \\

\midrule

\multirow{4}{*}{DeepSeek-Coder}
& Random Partition      & \textbf{31.9} & 33.0 & 34.8 & 33.0 & 22.9 & 23.9 & 24.1 & 25.1\\
& Interleaved Partition & 31.2 & \textbf{35.2} & \textbf{36.9} & 33.5 & 23.9 & 24.7 & 25.4 & \textbf{26.3} \\
& Clustering Partition  & 30.4 & 32.6 & 33.5 & 35.6 & 23.9 & 23.9& 24.4 & 25.9 \\
& Sequential Partition  & 31.7 & \textbf{35.2} & 32.6 & \textbf{36.5} & \textbf{24.1} & \textbf{25.5} & \textbf{26.3} & \textbf{26.3} \\

\bottomrule
\end{tabular}
\label{tab:partition_strategy}
\end{table}

\begin{table}[t]
\centering
\small
\setlength{\abovecaptionskip}{0.1cm}
\caption{Generation time comparison of different draft selection strategies in the Reduce phase. The text in bold indicates the shortest generation time.}
\begin{tabular}{llcccccccc}
\toprule
\multirow{2}{*}{Model} & \multirow{2}{*}{Draft Selection Strategy} 
& \multicolumn{4}{c}{CoderEval} 
& \multicolumn{4}{c}{DevEval} \\
\cmidrule(lr){3-6} \cmidrule(lr){7-10}
& & K=3 & K=5 & K=7 & K=10 & K=3 & K=5 & K=7 & K=10 \\
\midrule

\multirow{2}{*}{Qwen2.5-Coder}
& Random Draft      & \textbf{1.9} & 2.3 & 2.8 & \textbf{2.9}  &  \textbf{3.6} & \textbf{3.7} & \textbf{4.0} & 4.7 \\
& First Draft       & \textbf{1.9} & \textbf{2.2} & \textbf{2.5} & \textbf{2.9}  & \textbf{3.6} & \textbf{3.7} & 4.0 & \textbf{4.6}  \\

\midrule

\multirow{2}{*}{DeepSeek-Coder}
& Random Draft      & \textbf{4.5} & 5.8 & 6.3 & \textbf{6.4} & \textbf{5.7 }& \textbf{5.7} & 6.4 & 6.3 \\
& First Draft       & \textbf{4.5} & \textbf{5.6} & \textbf{5.7} & {6.7} & \textbf{5.7} & \textbf{5.7} & \textbf{5.8} & \textbf{6.1}  \\

\bottomrule
\end{tabular}
\label{tab:draft_selection_time}
\end{table}
To investigate the impact of different context partitioning strategies in the \textit{Map phase} on code generation quality, we conduct additional experiments incorporating the following four representative partitioning approaches, enabling a systematic comparison of their effects on generation performance.
% . We consider four representative context partition strategies:

\begin{itemize}
    \item \textbf{Sequential Partition}: The partition strategy used in our paper. Given the retrieved context blocks ordered by their similarity to the query, this strategy directly splits them into consecutive groups of a fixed size. Specifically, the first $m$ blocks form the first group, the next $m$ blocks form the second group, and so on. Consequently, this strategy preserves the original retrieval ranking and places highly ranked contexts into earlier groups, making the first group the most relevant one.

    \item \textbf{Random Partition}: This strategy first randomly shuffles all retrieved context blocks using a fixed random seed, and then applies the same consecutive grouping procedure as Sequential Partition. The purpose is to break the original relevance order and examine whether the grouping strategy is sensitive to the relevant ranking of retrieved contexts.

    \item \textbf{Interleaved Partition}: This strategy first determines the number of groups based on the total number of retrieved blocks and the group size, and then distributes contexts into groups in a round-robin manner. For example, if there are $n$ groups, the first group contains blocks at positions $1, 1+n, 1+2n,\dots$, whilst the second group contains blocks at positions $2, 2+n, 2+2n,\dots$. In this way, each group mixes contexts from different relevance levels.

    \item \textbf{Clustering Partition}: This strategy groups retrieved context blocks according to their vector representations. Specifically, each block is first transformed into a feature vector, and K-Means clustering is then applied to divide all blocks into several clusters, where the number of clusters is determined by the expected group size. This strategy aims to group semantically similar code snippets together, but it introduces additional clustering overhead. 
\end{itemize}

As shown in Table \ref{tab:partition_strategy}, different context partition strategies lead to overall comparable performance across both models and benchmarks, and no single strategy consistently dominates across all settings. While Random and Interleaved Partition occasionally achieve higher Pass@1 scores at specific values of $K$ (e.g., at larger $K$ on CoderEval), these improvements are not stable and often do not generalize across datasets or models. Similarly, Clustering-based Partition shows competitive performance in some cases, but fails to provide consistent gains despite its additional computational overhead.

These observations suggest that the partition strategy itself is not a primary bottleneck in the Map phase. Instead, as long as each group contains a reasonable subset of retrieved contexts, the subsequent draft generation and selection mechanism can effectively filter useful information. In this sense, different partition strategies mainly affect how context diversity and relevance are distributed across groups, but their overall impact on final generation quality remains limited. Among all strategies, Sequential Partition exhibits relatively stable performance across most configurations. By preserving the original retrieval order, it groups highly relevant contexts together, which helps the draft model generate more coherent and focused drafts. At the same time, it avoids introducing additional randomness or structural perturbations, leading to more consistent behavior.

\textbf{Practical Recommendations.}
In practice, Sequential Partition is recommended as the default choice due to its simplicity, efficiency, and stable performance. Interleaved Partition may be considered when relevant contexts are unevenly distributed and increased diversity across groups is desired, but its benefits are not consistent. Random Partition generally introduces unnecessary variance, and Clustering-based Partition is less attractive due to its additional preprocessing cost without clear performance advantages. Overall, Sequential Partition is sufficient and effective for the \textit{Map phase} and we used in our paper.

\subsection{Impact of Draft Selection Strategy in Reduce Phase}
\label{diss:draft_selection}

To further investigate the impact of draft selection in the \textit{Reduce phase}, we compare different strategies for choosing the draft used in parallel verification. Specifically, we consider two representative strategies:

\begin{itemize}
\item \textbf{First Draft}: Select the first draft generated from the most relevant context group (i.e., the group with highest similarity to the query). This strategy is adopted as the default choice in our method. Since the context groups are constructed in descending order of relevance, the first draft is more likely to align with the target model’s token distribution, leading to a higher acceptance rate during parallel verification and improved decoding efficiency.

\item \textbf{Random Draft}: Randomly select a draft from all candidate drafts generated in the \textit{Map phase}. This strategy is introduced as a control setting to evaluate the sensitivity of the \textit{Reduce phase} to draft quality. By removing the relevance-based selection bias, Random Draft allows us to assess how the impact of draft quality (the alignment with the target LLM's answer) on generation efficiency
\end{itemize}

As shown in Table \ref{tab:draft_selection_time}, different draft selection strategies lead to overall comparable generation time across both models and benchmarks. But selecting the first draft exhibits more stable and consistently competitive performance across different configurations. This can be attributed to the quality of the selected draft. The first draft is generated from the most relevant context group, making it more likely to align with the target model’s token distribution. As a result, a larger proportion of draft tokens can be directly accepted during parallel verification, reducing the need for fallback decoding and improving overall efficiency. In contrast, Random Draft may select drafts derived from less relevant context groups, which are more likely to deviate from the target model’s predictions. This leads to increased token rejections and additional validation overhead, slightly degrading decoding efficiency.

\textbf{Practical Recommendations.}
In practice, selecting the first draft is recommended as the default strategy, as it consistently provides high-quality drafts that better match the target model’s predictions. Given its simplicity and effectiveness, a deterministic and relevance-based draft selection strategy is sufficient for achieving efficient decoding in the \textit{Reduce phase}.
\section{Threats to Validity}
\textbf{Threats to External Validity.}
First, our experiments are conducted on two widely used repository-level code generation benchmarks, namely CoderEval and DevEval, which mainly focus on Python repositories. Although these benchmarks may not fully represent other programming languages, they cover diverse real-world scenarios and are widely adopted in existing code generation studies~\cite{EmpiricalRAGCode,iterative,A3CodGen,reposcope,fastcoder}. Second, we evaluate \ourmodel{} using two representative open-source LLM families, Qwen2.5-Coder and DeepSeek-Coder. We adopt open-source models because our method requires access to token-level probability distributions (e.g., for parallel verification), which is typically unavailable in commercial APIs, and because they allow full control over inference to ensure stable efficiency measurements. Due to resource constraints, we do not include extremely large models. However, our framework is model-agnostic and does not rely on specific architectures, and thus can be extended to larger or proprietary models when decoding signals are accessible.
Third, we adopt BM25 as the retrieval method. Although \ourmodel{} is designed to be retrieval-agnostic, different retrieval strategies (e.g., dense or graph-based retrieval) may interact differently with our selection mechanism.

\textbf{Threats to Internal Validity.}
Internal validity concerns the correctness of the experimental design and implementation. To ensure fair comparisons, we carefully follow the original papers for all baseline methods and adopt official or widely used implementations. We align hyperparameters, preprocessing, and training settings as closely as possible, though minor differences (e.g., for RL-Coder) may remain. Second, the experimental environment (e.g., GPU hardware and system dependencies) may influence efficiency measurements such as inference time; we mitigate this by running all experiments on the same hardware. Finally, using small models for draft generation may be limited by their capacity, potentially overlooking relevant context. While our results demonstrate the overall effectiveness of \ourmodel{}, this limitation will be addressed in future work.

% \textbf{Threats to Construct Validity.}
% Construct validity concerns whether the evaluation metrics accurately reflect the intended objectives. We use Pass@K (primarily Pass@1) to measure functional correctness via test cases. However, this metric does not capture other aspects of code quality, such as readability or maintainability. Efficiency is measured by inference time and token consumption, which reflect computational cost but may not fully capture real-world user experience (e.g., interactive latency). In addition, context selection quality is evaluated indirectly through downstream generation performance rather than explicit relevance metrics. While this aligns with end-task objectives, it may not fully isolate the contribution of context selection. Finally, the use of static analysis tools (e.g., Tree-sitter) to extract API calls may introduce approximation errors.

\section{Conclusion}
In this paper, we propose \ourmodel{}, an efficient context selection framework for repository-level code generation. By adopting a Map–Reduce paradigm with draft-guided context selection, \ourmodel{} effectively filters noisy and redundant contexts while preserving informative contexts. The proposed SADGS mechanism integrates API call relationships and logical similarity to identify relevant context, and the parallel verification strategy further improves generation efficiency. Extensive experiments on CoderEval and DevEval demonstrate that \ourmodel{} consistently improves generation quality while significantly reducing token consumption and inference time, achieving a better balance between effectiveness and efficiency compared to existing methods.
In future work, we plan to extend \ourmodel{} to larger models and more diverse programming environments.
\bibliographystyle{ACM-Reference-Format}
\bibliography{sample-base}

% \appendix









\end{document}